\documentclass[11pt]{amsart}
\usepackage{graphicx}
\usepackage{amscd}
\usepackage{amsmath}
\usepackage{amsfonts}
\usepackage{amssymb}
\usepackage{bbm}
\usepackage{setspace}
\usepackage{enumerate}         
\usepackage{color}
\usepackage{url}
\usepackage{amsthm}
\usepackage{hyperref}
\usepackage{bm}
\usepackage{xy}
\usepackage{color}
\usepackage{ulem}
\usepackage{cancel}
\usepackage{subfig}
\usepackage{bbm}
\usepackage{etoolbox}
\usepackage[pagewise]{lineno}
\usepackage{mathrsfs}
\usepackage{algpseudocode}
\usepackage{algorithmicx}
\usepackage[ruled,vlined]{algorithm2e}
\usepackage{neuralnetwork}

\allowdisplaybreaks[4]

\usepackage{natbib}

\usepackage{geometry}
\usepackage{verbatim}

\usepackage{tikz}
\usetikzlibrary{positioning,shapes,shadows,arrows}
\tikzstyle{abstract}=[rectangle, draw=black, rounded corners, fill=blue!40, drop shadow,
        text centered, anchor=north, text=white, text width=2.5cm]
\tikzstyle{comment}=[rectangle, draw=black, rounded corners, fill=green, drop shadow,
        text centered, anchor=north, text=white, text width=2cm]
\tikzstyle{myarrow}=[->, >=open triangle 90, thick]
\tikzstyle{line}=[-, thick]

\theoremstyle{plain}
\newtheorem{theorem}{Theorem}[section]

\theoremstyle{remark}
\newtheorem{remark}[theorem]{Remark}

\theoremstyle{hp}
\newtheorem{hp}{Assumption}

\numberwithin{equation}{section}

\newcommand{\rsto}{]\!\kern-1.8pt ]}
\newcommand{\lsto}{[\!\kern-1.7pt [}

\numberwithin{equation}{section}

\renewcommand{\emph}[1]{\textit{#1}}
\newcommand{\Ind}[1]{\mathbbm{1}_{\left\{#1\right\}}}

\newcommand{\FF}{\mathbb{F}}
\newcommand{\GG}{\mathbb{G}}
\newcommand{\HH}{\mathbb{H}}
\newcommand{\RR}{\mathbb{R}}
\newcommand{\QQ}{\mathbb{Q}}

\newcommand{\NN}{\mathbb{N}}
\newcommand{\EE}{\mathbb{E}}

\newcommand{\cF}{\mathcal{F}}
\newcommand{\cG}{\mathcal{G}}

\newcommand{\cI}{\mathcal{I}}

\newcommand{\cL}{\mathcal{L}}

\newcommand{\cV}{\mathcal{V}}
\newcommand{\cX}{\mathcal{X}}

\newcommand{\cZ}{\mathcal{Z}}

\newcommand{\Ex}[2]{\mathbb{E}^{#1}\left[#2\right]}                     
\newcommand{\Excond}[3]{\mathbb{E}^{#1}\left[\left.#2\right|#3\right]}  

\newcommand{\diag}{\mathop{\mathrm{diag}}}

\renewcommand{\cite}{\citet}

\makeatletter
\patchcmd{\@maketitle}
  {\ifx\@empty\@dedicatory}
  {\ifx\@empty\@date \else {\vskip3ex \centering\footnotesize\@date\par\vskip1ex}\fi
   \ifx\@empty\@dedicatory}
  {}{}
\patchcmd{\@adminfootnotes}
  {\ifx\@empty\@date\else \@footnotetext{\@setdate}\fi}
  {}{}{}
\newcommand{\subjclassname@JEL}{JEL Classification}
\makeatother

\begin{document}

\title[Deep xVA solver]{Deep xVA solver -- A neural network based  counterparty credit risk management framework}

\author{Alessandro Gnoatto}
\address[Alessandro Gnoatto]{University of Verona, Department of Economics, \newline
\indent via Cantarane 24, 37129 Verona, Italy}
\email[Alessandro Gnoatto]{alessandro.gnoatto@univr.it}

\author{Athena Picarelli}
\address[Athena Picarelli]{University of Verona, Department of Economics, \newline
\indent via Cantarane 24, 37129 Verona, Italy}
\email[Athena Picarelli]{athena.picarelli@univr.it}%

\author{Christoph Reisinger}
\address[Christoph Reisinger]{Oxford University, Mathematical Institute \newline
\indent ROQ, Woodstock Rd, Oxford, OX2 6GG, UK}
\email[Christoph Reisinger]{christoph.reisinger@maths.ox.ac.uk}%

\begin{abstract}
In this paper, we present a novel computational framework for portfolio-wide risk management problems, where
the presence of a potentially large number of risk factors makes traditional numerical techniques ineffective.
The new method utilises a coupled system of BSDEs for the valuation adjustments (xVA) and solves these by a recursive application of a neural network based BSDE solver.
This not only makes the computation of xVA for high-dimensional problems feasible, but also produces hedge ratios and dynamic risk measures for xVA, and allows simulations of the collateral account.
\end{abstract}

\keywords{CVA, DVA, FVA, ColVA, xVA, EPE, Collateral, xVA hedging, Deep BSDE Solver, Neural Networks}
\subjclass[2010]{91G60, 91G20, 91G40. \textit{JEL Classification} G13, G17}

\date{\today}

\maketitle

\section{Introduction}\label{intro}
As a consequence of the 2007--2009 financial crisis, academics and practitioners have been redefining and augmenting key concepts of risk management. This made it necessary to reconsider many widely used methodologies in quantitative and computational finance.

It is now generally accepted that a reliable valuation of a financial product should account for the possibility of default of any agent involved in the transaction. Moreover, the trading activity is nowadays funded by resorting to different sources of liquidity (the interest rate multi-curve phenomenon; see, e.g., \cite{CFGaffine}), so that the existence of a single funding stream with a unique risk-free interest rate no longer represents a realistic assumption.
Additionally, the increasingly important role of collateral agreements demands for a portfolio-wide view of valuation.

These stylized facts are incorporated into the valuation equations through value adjustments (xVA). Value adjustments are terms to be added to, or subtracted from, an idealized reference portfolio value, computed in the absence of frictions, in order to obtain the final value of the transaction. 

The literature on counterparty credit risk and funding is large and we only attempt to provide insights on the main references as they relate to our work. Possibly the first contribution on the subject is a model for credit risk asymmetry in swap contracts in \cite{duhu96}. Before the 2007--2009 financial crisis, we have the works of \cite{brigoMasetti} and \cite{cherubini05}, where the concept of credit valuation adjustment (CVA) is analyzed. The possibility of default of both counterparties involved in the transaction, represented by the introduction of the debt valuation adjustment (DVA), is investigated, among others, in \cite{bripapa11, bricapa14}.

Another important source of concern to practitioners apart from default risk is represented by funding costs. A parallel stream of literature emerged during and after the financial crisis to generalize valuation equations in the presence of collateralization agreements. In a Black-Scholes economy, \cite{pit10} gives valuation formulas both in the collateralized and uncollateralized case. Generalizations to the case of a multi-currency economy can be found in \cite{pit12},   \cite{fushita10b, fushita09}, and \cite{gs2020}.
The funding valuation adjustment (FVA) is derived under alternative assumptions on the Credit Support Annex (CSA) in \cite{papebri2011}, while \cite{bripa2014ccp} also discusses the role of central counterparties for funding costs. A general approach to funding in a semimartingale setting is provided by \cite{BieRut15}.

Funding and default risk need to be united in a single risk management framework to account for all possible frictions and their interplay. Contributions in this sense can be found in \cite{bbfpr2018} by means of the so-called discounting approach. {  \cite{bj2013, bj2011} generalize the classical Black-Scholes replication approach to include some of the aforementioned effects. 
A more general backward stochastic differential equation (BSDE) approach is provided by \cite{crepey2015a, crepey2015b} and
\cite{BiCaStu2018, BiCaStu2019}. The equivalence between the discounting approach and the BSDE-based replication approaches is demonstrated in \cite{bbfpr2018}. 


A common fundamental feature of such generalized risk management frameworks is the necessity to adopt a portfolio-wide point of view in order to properly account for risk mitigation benefits arising from diversified positions.
Adopting such portfolio-wide models, as is the present market practice in financial institutions, involves high-dimensional joint simulations of all positions within a portfolio.  

Commonly used numerical techniques (see for instance \cite{Shoftner08,KarJaiOos16,BroDuMoa15,JosKwo16}) make use of regression approaches, based on a modification of the Least-Squares Monte Carlo approach in  \cite{LonSch01},  to alleviate the high computational cost of fully nested Monte Carlo simulations such as those initially proposed in \cite{GorJun10,BroDuMoa11}. 
We refer to \cite{alcacre2017} for a high-performance GPU implementation of nested Monte Carlo for bilateral xVA computations in a modern set-up including credit, margin and capital, for a large book of about 200,000 trades with 2000 counterparties.

 {For an application of adjoint algorithmic differentiation (AAD) to xVA simulation by regression see, for instance, \cite{capriotti2017aad, Fries19aad}.}



An alternative, hybrid, approach to counterparty risk computations is taken in \cite{GraFenKanOos14}, where standard pricing methods are applied to the products in the portfolio and outer Monte Carlo estimators for exposures.
Techniques based purely on PDEs generally suffer from the \emph{curse of dimensionality}, a rapid increase of computational cost in presence of high dimensional problems.
A PDE approach with factor-based dimension reduction has been proposed in  \cite{GraKanRei18}.



 
%

In the broader context of high-dimensional problems involving large amounts of data,
machine learning techniques have witnessed dramatically increasing popularity.
Of particular interest is the concept of an artificial neural network (ANN). From a mathematical perspective, ANNs are multiple nested compositions of
relatively simple multivariate functions. 
The term deep neural networks refers to ANNs with several interconnected layers.
One remarkable property of ANNs is given in the `Universal Approximation Theorem', which essentially states that any continuous function in any dimension can be represented to arbitrary accuracy by means of an ANN, and has been proven in different versions,
starting from the remarkable insight of Kolmogorov's Representation Theorem in \cite{kolmo56} and the seminal works of \cite{cybenko89} and \cite{hornik91}.
Recently, building heavily on earlier work of \cite{jentzen2018proof},
the recent results by \cite{reizha19} have proven that deep ANNs can overcome the curse of dimensionality for
approximating (nonsmooth) solutions of partial differential equations arising from (open-loop control of) SDEs.
A result to the same effect has been shown for heat equations with a zero-order nonlinearity in \cite{hutzenthaler2018overcoming}. This is potentially useful
in the context of risk management as simple models for CVA can be expressed in this form. 
For a recent literature survey of applications of neural networks to pricing, hedging and risk management problems more generally we refer the reader to \cite{ruf2019neural}.


In this paper, we investigate the application of ANNs to solve high-dimensional BSDEs arising from risk management problems.
Indeed, in the classical continuous-time mathematical finance literature the random behavior of the simple financial assets composing a  portfolio is typically described by means of multi-dimensional Brownian motions and forward stochastic differential equations (SDEs). In  this setting, BSDEs naturally arise as a representation of the evolution of the hedging portfolio, where the terminal condition represents the target payoff (see, e.g., \cite{elkpenque97}). In essence, (numerically) solving a BSDE is equivalent to identifying a risk management strategy.

Numerical BSDE methods published recently for xVA computations for single derivatives include \cite{borovykh2018efficient}. 
The difficulty of extending these computational techniques to the portfolio setting is alluded to in Remark 11 of \cite{ninomiya2019higher}.

Here, we will consider a discretized version of the BSDE and parametrize the (high dimensional) control (i.e., hedging) process at every point in time by means of a family of ANNs. 
Once written in this form, BSDEs can be viewed as model-based reinforcement learning problems. The ANN parameters are then
fitted so as to minimize a prescribed loss function. 

The line of computational methods we follow has been initiated in the context of high-dimensional nonlinear PDEs in \cite{ehanjen17} and further investigated
in \cite{HanLon18} and \cite{fujtaktak19}, and has led to a class of methods for the solution of BSDEs (characterised by parametrisation of the Markovian control by ANNs), which we will collectively refer to as the Deep BSDE Solver for simplicity.
 By way of financial applications, and xVA specifically,
a primal-dual extension to the Deep BSDE Solver has been developed in \cite{henlab17} and tested on stylised CVA- and IM(Initial Margin)-type PDEs;
the Deep BSDE Solver has also been applied specifically to exposure computations for a Bermudan swaption and a cross-currency swap in \cite{SheGre18}.

Our approach goes beyond these earlier works in the following regards: we
\begin{itemize}
\item consider a rigorous, generic BSDE model for the dynamics of xVA, including CVA, DVA, FVA and ColVA (collateral valuation adjustment), for a derivative portfolio;
\item
introduce algorithms for the computation of `non-recursive' xVAs -- such as CVA and DVA -- and `recursive' xVAs -- such as FVA -- by (recursive) application of the Deep BSDE Solver, and deduce \emph{a posteriori} bounds on the error of the neural network approximations;
\item
show how the method can be used for the simulation of xVA sensitivities and collateral, and provide careful numerical tests, showing good (i.e., basis point) accuracy for different adjustment computations, including an example with 100 underlying assets.
\end{itemize}

We will refer to our method as \emph{Deep  xVA Solver}.
More recently, conditional risk measure computations (VaR and ES), based on deep learning regression,
have been proposed in an xVA framework in \cite{ACHS20}, using a similar numerical approach to the one developed independently for BSDEs in \cite{HPW20}.
Different from \cite{ehanjen17}, this solver approximates the value function, not the control, by means of an ANN and reconstructs it at each time step by dynamic programming techniques. 
A comparison of the performance and robustness of the two approaches will require comprehensive testing in industry-relevant settings. 
We see as a structural advantage of our algorithm that it allows to obtain the xVA hedging strategy with no need of further computation (i.e. differentiation).

The applicability of the presented methodology is largely independent of the particular choice of the xVA framework.
In particular, we do not take a position in the so-called \textit{FVA debate} or on the question of including KVA in the pricing equation.
The term \textit{FVA debate} here refers to the possible overlap between the debt value adjustment (DVA) and the funding benefit adjustment (FBA). This overlap has been addressed 
in \cite{bfp2019}. The inclusion of KVA is still debated, noting, for example, the recent criticism of KVA in \cite{ads2019}.
Our approach is general enough to accommodate different specifications of the price decomposition. 
In particular, our methodology can be applied immediately to the framework of \cite{bfp2019}.

We restrict the presentation of the method to a single counter-party -- or `netting set' -- for simplicity, as is routinely done in banks.
There are economic grounds for extending the computation to multiple netting sets simultaneously  (see, e.g., \cite{ACHS20}) and our method 
generalises accordingly.

The paper is organized as follows. The financial framework is established in Section \ref{Sect:preliminaries}. In Section \ref{sec:algo}, after shortly recalling the main features of the Deep BSDE Solver presented in \cite{ehanjen17}, the algorithm for xVA computation is introduced. Numerical results for a selection of test cases are shown in 
Section \ref{sec:numerics}, while Section \ref{sec:concl} concludes.

\section{The financial market}\label{Sect:preliminaries}

{
For concreteness, we adopt the market setup of \cite{bgo2019} and subsequently formulate our computational methods in the context of this model.
Let us re-iterate the point elaborated in the introduction, however, that the computational framework, which is the focus of this article, is adaptable to a range of model specifications.
}

We fix a time horizon $T<\infty$ for the trading activity of two agents named the \textit{bank} (B) and the 
\textit{counterparty} (C). 
Unless otherwise stated, throughout the paper we assume the bank's perspective and refer to the bank as the \textit{hedger.} 

All underlying processes are modeled over a probability space $\left(\Omega,\cG,\GG,\QQ\right),$ where $\GG=\left(\cG_t\right)_{t\,\in\,[0,T]} 
\subseteq \cG$ is a filtration satisfying the usual assumptions ($\cG_0$ is assumed to be trivial).
We denote by $\tau^B$ and $\tau^C$ the \textit{time of default} of the bank and the counterparty, respectively. 
Specifically, we assume that $\GG=\FF\,\vee\,\HH,$ where $\FF = (\cF_t)_{t\,\in\,[0,T]}$ is a reference filtration satisfying the usual assumptions 
and $\HH=\HH^B\vee \HH^C,$ with $\HH^j=\big(\mathcal{H}^j_t\big)_{t\,\in\,[0,T]}$ for $\mathcal{H}^j_t 
= \sigma\big(\left.H_u\right|u\leq t\big)$,  and $H^j_t:=\Ind{\tau^j\leq t}, \, j\,\in\, \{B,C\}.$   
We set 
$$
\tau = \tau^C \wedge \tau^B. 
$$

In the present paper we will extensively make use of the so called \textit{Immersion Hypothesis} (see, e.g., \cite{bielecki2004credit}).

\begin{hp} \label{hp:H}   
Any local $(\FF,\QQ)$-martingale is a local $(\GG,\QQ)$-martingale. 
\end{hp}

We consider  the following spaces:
\begin{itemize}
\item $L^2(\RR^d)$ is the space of all $\cF_T$-measurable $\RR^d$-valued random variables $X:\Omega\mapsto\RR^d$ such that $\left\| X\right\|^2=\Ex{}{\left|X\right|^2}<\infty$.
\item $\HH^{2,q\times d}$ is the space of all predictable $\RR^{q\times d}$-valued processes $\phi:\Omega\times [0,T]\mapsto \RR^{q\times d}$ such that $\Ex{}{\int_0^T|\phi_t|^2dt}<\infty$.
\item $\mathbb{S}^2$ the space of all adapted processes $\phi:\Omega\times [0,T]\mapsto \RR^{q\times d}$ such that $\Ex{}{\sup_{0\leq t\leq T}|\phi_t|^2}<\infty$.
\end{itemize}

\subsection{Basic traded assets} 

\subsubsection*{Risky assets} \label{risky_asset}
For $d \geq 1,$ we denote by $S^i$, $i=1,\ldots, d$, the \textit{ex-dividend price} (i.e. the price) of risky securities. 
All $S^i$ are assumed to be c\`adl\`ag $\FF$-semimartingales. 


Let $W^\QQ=\big(W^\QQ_t\big)_{t\,\in\,[0,T]}$ be a $d$-dimensional $(\FF,\QQ)$-Brownian motion (hence a $(\GG,\QQ)$-Brownian motion, 
thanks to Assumption~\ref{hp:H}). We introduce the coefficient functions
$\mu: \; \RR_{+}\times \RR^d\mapsto \RR^d$,
$\sigma: \; \RR_{+}\times \RR^d \mapsto \RR^{d\times d}$,
which are assumed to satisfy standard conditions ensuring existence and uniqueness of strong solutions of SDEs driven by the Brownian motion $W^\QQ$. 
We assume that
\begin{align}\label{eq:asset_prox}
\begin{cases}
\mathrm d S_t \!\!\!\! &=\mu(t,S_t) \, \mathrm  d t+\sigma(t,S_t) \, \mathrm d W^\QQ_t,\\
S_0 \!\!\!\! &=s_0\in\RR^d,
\end{cases}
\end{align}
on $[0,T]$, {where $S_t =(S^1_t, \ldots, S^d_t)\in \RR^d$.}
Note that we are not postulating that the processes $S^i$ are positive. 


Throughout the paper we assume that the market is complete for the sake of simplicity. 

\subsubsection*{Cash accounts} \label{cash_account}

Given a stochastic return process 
$x := (x_t)_{t \geq 0}$, which is assumed bounded, right-continuous and $\FF$-adapted, we define
the cash account $B^x$ with unitary value at time $0$, as the strictly positive 
continuous processes of finite variation 
\begin{align} \label{def:cash_account}
B^x_t := \exp\left\{\int_0^t {x}_s \, \mathrm d s\right\}, \,t \,\in\, [0,T]. 
\end{align}
In particular, $B^x := (B^x_t)_{t\,\in\,[0,T]}$ is also continuous and adapted.

\subsubsection*{Defaultable bonds} \label{def_bond}

Default times are assumed to be exponentially distributed random variables with time-dependent intensity 
$$\Gamma^j_t=\int_0^t\lambda^{j,\QQ}_s \, \mathrm d s, \quad t \,\in\, [0,T], \, \, j\,\in\,\{B,C\},$$ 
where $\lambda^{j,\QQ}$ are non-negative bounded processes.

We introduce two risky bonds with maturity $T^\star \leq T$ issued by 
the bank and the counterparty. We directly state their dynamics under $\QQ$. We refer to \cite{bgo2019} for more details. The risky bonds evolve according to

\begin{align}\label{eq:bond_prox}
\mathrm d P^j_t=r^j_tP^j_t \, \mathrm d t-P^j_{t-}\mathrm \, d M^{j,\QQ}_t, \quad j\in\{B,C\},
\end{align}
where $M^{j,\QQ}$, $j\in\{B,C\}$ are  compensated Poisson random measures, see equation (3.6) in \cite{bgo2019}.

\subsection{xVA framework}\label{Sect:single}

We consider a family of contingent claims within a portfolio with agreed dividend stream $A^m=\left(A^m_t\right)_{t\in[0,T] }$, $m=1,\ldots, M$, and set $\bar{A}^m_t :=\Ind{t<\tau}A^m_t+\Ind{t\geq \tau}A^m_{\tau-}$.  {We let  $T_m\leq T$  denote the maturity time of the $m$-th contract.}
The value of the single claim within the portfolio, ignoring any  counterparty risk or funding issue, that we refer to as \emph{clean values}, are denoted by $(\widehat{V}^m_t)_{m=1,\ldots, M}$   and satisfy the following forward-backward stochastic differential equations (FBSDEs), for $m=1,\ldots,M$,
\begin{align} \label{cleanBSDE}
\begin{cases}
-\mathrm d\widehat{V}^m_t = \mathrm d A^m_t-r_t\widehat{V}^m_t \, \mathrm dt -\sum_{k=1}^d\widehat{Z}^{m,k}_t \, \mathrm d W^{k,\QQ}_t, \\
\widehat{V}^m_{T_{m} } = 0,
\end{cases}
\end{align}
which reads, in integral form,
\begin{align}\label{fair_price:clean_mkt}
\widehat{V}^m_t:=\Excond{\QQ}{B^r_t\int_{(t,T_m]} \frac{\mathrm dA^m_u}{B^r_u}}{\cF_t}, \quad  
t \,\in\, [0,T_m],
\end{align}
where $r$ is a collateral rate in an idealized perfect collateral agreement.

For simplicity, we restrict ourselves to Europan-type contracts and, {for  $m=1,\ldots M$, denoted by the Lipschitz function $g_m$ the payoff of the option we write $A^m_t=\Ind{t\geq T_m}g_m(S_{T_m})$}.  {In this case,  with an abuse of notation, instead of equation  \eqref{cleanBSDE} we consider  } 
\begin{align} \label{cleanBSDEeu}
\begin{cases}
-\mathrm  d\widehat{V}^m_t = -r_t\widehat{V}^m_t \, \mathrm  dt -\sum_{k=1}^d\widehat{Z}^{m,k}_t \, \mathrm dW^{k,\QQ}_t, \\
\widehat{V}^m_{T_m} = g_m(S_{T_m}).
\end{cases}
\end{align}
Observe that the system  \eqref{eq:asset_prox} and \eqref{cleanBSDEeu}  is decoupled, in the sense that the forward equation \eqref{eq:asset_prox} does not exhibit a dependence on the backward component. \\
We continue to follow the framework of \cite{bgo2019}, where the portfolio dynamics are stated in the form of a BSDE under the enlarged filtration $\GG$. 
We set
\begin{subequations}
\begin{align} 
\label{eq:Zk}
Z^k_t & :=\sum_{i=1}^d\xi^i_t\sigma^{i,k}(t,S_t), & k=1,\ldots, d, \\ \label{eq:Uj}
U^j_t & :=-\xi^j_tP^j_{t-},& j\in \{B,C\}, \\ \label{eq:driver}
f(t,V,C) & :=-\left[(r^{f,l}_t-r_t)\left(V_t-C_t\right)^+-(r^{f,b}_t-r_t)\left(V_t-C_t\right)^-
\right.\\ \nonumber
&\left.
+(r^{c,l}_t-r_t)C^+_t-(r^{c,b}_t-r_t)C^-_t\right],
\end{align}
\end{subequations}
where 
\begin{itemize}
\item $\xi^i$, $i=1,\ldots, d$, are the positions in risky assets, while $\xi^B,\xi^C$ are the position in the bank and counterparty bond respectively; 
\item $r^{f,l},r^{f,b}$ represent unsecured funding lending and borrowing rates;
\item $r^{c,l},r^{c,b}$ denote the interest on posted and received variation margin (collateral);
\item {$\sigma^{i,k}(t,S_t)$ is the $(i,k)$-th entry  of the matrix $\sigma(t,S_t)$, for $i,k=1,\ldots, d$;}
\item $C^+$ and $C^-$ represent the posted and received variation margin/collateral and $C=C^+-C^-$.
\end{itemize}

All above processes are assumed to satisfy suitable regularity conditions ensuring existence and uniqueness for a solution  to  BSDE \eqref{eq:GBSDEdiff} below. 
Both posted and received collateral are assumed to be Lipschitz functions of the clean value of the derivative portfolio and we will write $C_t=C(V_t)$.

We denote by $V$ the \emph{full contract} value, i.e. the portfolio value including counterparty risk and multiple curves.  The  $\GG$-BSDE for the portfolio's dynamics then has the form on $\{t<\tau\}$
\begin{align} \label{eq:GBSDEdiff}
\begin{cases}
-\mathrm dV_t = \sum_{m=1}^M \mathrm d\bar{A}^m_t + \left(f(t,V,C) - r_t V_t\right) \, \mathrm dt 
-\sum_{k=1}^dZ^k_t \, \mathrm d W^{k,\QQ}_t - \sum_{j\in\{B,C\}}U^j_t \, \mathrm d M^{j,\QQ}_t, \\
V_\tau = \theta_\tau(\widehat{V},C), \qquad \text{with }\\
\theta_\tau( \widehat{V},C):=\widehat{V}_\tau
+\Ind{\tau^C< \tau^B}(1-R^C)\left( \widehat{V}_\tau-C_{\tau-} 
\right)^--\Ind{\tau^B< \tau^C}(1-R^B)\left( \widehat{V}_\tau-C_{\tau-}\right)^+,
\end{cases}
\end{align}
 where $\widehat V_t := \sum_{m=1}^M\widehat{V}^m_t$ and $R^B$, $R^C$ are two positive constants representing the recovery rate of the bank and the counterparty, respectively.

In their Theorem 3.16, \cite{bgo2019} show that there exists a unique solution $(V,Z,U)$ for the $\GG$-BSDE \eqref{eq:GBSDEdiff}, 
and the process $V$ assumes the following form on $\{t<\tau \}$:
\begin{align}\label{eq:GBSDEint} 
   V_t = B^r_t\Excond{\QQ}{\sum_{m=1}^M\int_{(t,\tau \wedge T]}\frac{ \mathrm d\bar{A}^m_u}{B^r_u} 
		+ \int_t^{\tau \wedge T}\frac{f(u,V,C)}{B^r_u} \, \mathrm d u + \Ind{\tau\leq T}\frac{\theta_\tau(\widehat{V},C)}{B^r_\tau}}{\cG_t}.
\end{align}

To prove existence and uniquencess for the $\GG$-BSDE, \cite{bgo2019} employ the technique introduced by \cite{crepey2015a} and reformulate the problem under the reduced filtration $\FF$. Stated in such a form, the problem is also more amenable to numerical computations.

We consider the following $\FF$-BSDE on $[0,T]$:
\begin{align} \label{eq:XVApreDef}
\begin{cases}
-\mathrm  d\overline{\rm XVA}_t = \bar{f}(t , \widehat V_t, \overline{\rm XVA}_t) \, \mathrm  dt-\sum_{k=1}^d\overline{Z}^k_t \, \mathrm dW^{k,\QQ}_t, \\
\overline{\rm XVA}_T = 0, 
\end{cases}
\end{align}
where
\begin{align}
\begin{aligned}
\bar{f}(t , \widehat V_t, \overline{\rm XVA}_t)&:=-(1-R^C)\left(\widehat{V}_t-C_{t}\right)^-\lambda^{C,\QQ}_t \\
&+(1-R^B)\left(\widehat{V}_t-C_{t}\right)^+\lambda^{B,\QQ}_t\\
&+(r^{f,l}_t-r_t)\left(\widehat{V}_t-\overline{\rm XVA}_t-C_t\right)^+-(r^{f,b}_t-r_t)\left(\widehat{V}_t-\overline{\rm XVA}_t-C_t\right)^-\\
&+(r^{c,l}_t-r_t)C^+_t-(r^{c,b}_t-r_t)C^-_t-(r_t+\lambda^{C,\QQ}_t+\lambda^{B,\QQ}_t)\overline{\rm XVA}_t .
\end{aligned}
\end{align}

By standard results on BSDEs, see e.g.\ \cite[Theorem~4.1.3, Theorem~3.1.1]{Delongbook}, 
the existence and uniqueness of solutions  $(\widehat V^m,\widehat Z^m) \in \mathbb{S}^{2}(\mathbb{R}) \times \mathbb{H}^{2,q\times 1}$, for $m=1,\ldots, M$, and $(\overline{\rm XVA}, \overline Z) \in \mathbb{S}^{2}(\mathbb{R}) \times \mathbb{H}^{2,q\times 1}$ to, respectively, \eqref{cleanBSDEeu} and \eqref{eq:XVApreDef}, holds under the following conditions:
\begin{align*}
&   r^{f,l}, r^{f,b}, r^{c,l}, r^{c,b}, r, \lambda^{B,\QQ}, \lambda^{C,\QQ} \text{ are bounded processes;}\\
&   |\mu(t,x)-\mu(t,x^{\prime})| +   |\sigma(t,x)-\sigma(t,x^{\prime})|  \leq C|x-x^{\prime}|, \\
&   |\sigma(t,x)| + |\mu(t,x)| \leq C(1+|x|).
\end{align*}

The process  $\overline{\rm XVA}$ coincides with the  \textit{pre-default}  xVA process. 
Indeed, given the pre-default value process $\overline{V}$ such that  $\overline{V}_t\Ind{t<\tau}=V_t\Ind{t<\tau}$, on $\{t<\tau\}$ the solution to \eqref{eq:GBSDEdiff} can be represented as
\begin{align*}
\overline{V}_t= \widehat V_t-\overline{\rm XVA}_t.
\end{align*}

%

Moreover, defining the process $\tilde{r}=\left(\tilde{r}_t\right)_{t\in[0,T]}$ 
as $\tilde{r}:=r+\lambda^{C,\QQ}+\lambda^{B,\QQ}$,
it has been shown in \cite[Corollary 3.17]{bgo2019} that the process $\overline{\rm XVA}$ admits the 
representation
\begin{align}\label{def:xvaPredef}
\overline{\rm XVA}_t = -\overline{\rm CVA}_t + \overline{\rm DVA}_t + \overline{\rm FVA}_t + \overline{\rm ColVA}_t,  
\end{align}
where 
\begin{align}
\overline{\rm CVA}_t& := B^{\tilde{r}}_t\Excond{\QQ}{(1-R^C)\int_t^{T}\frac{1}{B^{\tilde{r}}_u}\left(\widehat{V}_u-C_{u} \right)^-\lambda^{C,\QQ}_u \, \mathrm du}{\cF_t}, \label{eq:defCVA}\\
\overline{\rm DVA}_t& := B^{\tilde{r}}_t\Excond{\QQ}{(1-R^B)\int_t^{T}\frac{1}{B^{\tilde{r}}_u}\left(\widehat{V}_u-C_{u} 
\right)^+\lambda^{B,\QQ}_u \, \mathrm du}{\cF_t}, \label{eq:defDVA}\\
\overline{\rm FVA}_t& :=B^{\tilde{r}}_t \Excond{\QQ}{\int_t^{T}\frac{(r^{f,l}_u-r_u)\left(\widehat{V}_u-\overline{\rm XVA}_u-C_u\right)^+}{B^{\tilde{r}}_u} \, \mathrm du}{\cF_t} \label{eq:defFVA} \\
&\quad-B^{\tilde{r}}_t \Excond{\QQ}{\int_t^{T}\frac{(r^{f,b}_u-r_u)\left(\widehat{V}_u-\overline{\rm XVA}_u-C_u\right)^-}{B^{\tilde{r}}_u} \, \mathrm  du}{\cF_t},
\nonumber \\
\overline{\rm ColVA}_t& :=B^{\tilde{r}}_t \Excond{\QQ}{\int_t^{T}\frac{(r^{c,l}_u-r_u)C^+_u-(r^{c,b}_u-r_u)C^-_u}{B^{\tilde{r}}_u} \, \mathrm  du}{\cF_t}.\label{eq:colVA}
\end{align}

This representation highlights that the inclusion of different borrowing and lending rates introduces a non-zero funding adjustment which cannot be found independently of the other adjustments. {As a consequence, differently from CVA, DVA and ColVA, FVA presents a recursive structure\footnote{{The ColVA can be recursive or non-recursive depending on the specification of the collateralizaton agreement: if the collateral is a function of the clean value, then the ColVA term is non recursive (and this represent the situation typically found in practice) however, if the collateral depends on the whole value of the transaction, then also this term is recursive as the FVA}.}}. An algorithm to compute all valuations adjustments systematically in the `non-recursive' and `recursive' setting, especially with the view of potentially large portfolios, is the focus of the next sections.

\section{The algorithm}\label{sec:algo}
In this section, we introduce the algorithm for computing valuation adjustments by neural network approximations to the BSDE model from the previous section.
 We start by briefly recalling the main features of the Deep BSDE Solver in \cite{ehanjen17}.
 Then, we present the application of the solver to valuation adjustments and its extensions to obtain financially important quantities. 
 We first focus on  non-recursive adjustments, namely CVA and DVA, and then extend the approach to the recursive case (see the terminology introduced at the end of the last section).
 
In particular, we propose to use the Deep BSDE Solver in  \cite{ehanjen17} to approximate the dynamics of $\widehat V^m_u$, $m=1,\ldots, M$, $u\in [t,T]$,
which constitute the  portfolio $\widehat V_u=\sum^M_{i=1} \widehat V^m_u$. Once the portfolio value has been approximated and resulting collaterals computed, the value of the adjustment can be obtained either by inserting the values in an `outer' Monte Carlo computation for non-recursive adjustments, or applying a second time the Deep BSDE Solver to \eqref{eq:XVApreDef} in the recursive case. 

\subsection{The Deep BSDE Solver  of \cite{ehanjen17}} 
For the reader's convenience,
we describe in this section the main principles of the  algorithm in \cite{ehanjen17} as they are relevant to our setting. We consider a general FBSDE framework.
\\
Let $\left(\Omega,\cF,\QQ\right)$ be a probability space rich enough to support an $\RR^d$-valued Brownian motion $W^{\QQ}=(W^{\QQ}_t)_{t\in[0,T]}$. Let $\FF=(\cF_t)_{t\in[0,T]}$ be the filtration generated by $W^{\QQ}$, assumed to satisfy the standard assumptions. 
Let us consider an FBSDE in the following general form:
\begin{align}
X _ { t } &  = x + \int _ { 0 } ^ { t } b \left( s, X _ { s } \right) \mathrm d s + \int _ { 0 } ^ { t } a \left( s, X _ { s } \right)^\top \mathrm d W^{\QQ} _ { s } , \quad x \in \mathbb { R } ^ { d } \label{eq:forward}\\
Y_{t}     & = \vartheta (X_T) +\int_{t}^{T} h \left(s, X_{s}, Y_{s}, Z_{s}\right) \mathrm d s- \int_{t}^{T} Z_{s}^\top \mathrm d W^{\QQ}_{s}, \quad t \in[0, T], \label{eq:backward}
\end{align}
where the vector fields $b:[0,T]\times \mathbb{R}^d \mapsto \mathbb{R}^d$, $a:[0,T]\times \mathbb{R}^d\mapsto \mathbb{R}^{d\times d}$, $h:[0,T]\times \RR^d\times \RR \times \RR^d \mapsto \RR$ and $\vartheta:\RR^d\mapsto \RR$ satisfy
suitable assumptions ensuring existence and uniqueness results. We denote by  $(X^{x}_t)_{t\in[0,T]}\in \mathbb{S}^{2}(\mathbb{R}^d)$ and $(Y^y_t, Z_t)_{t\in [0,T]}\in \mathbb{S}^{2}(\mathbb{R}) \times \mathbb{H}^{2,q\times 1}$ the unique adapted  solution to \eqref{eq:forward} and \eqref{eq:backward}, respectively.  To alleviate notations, hereafter we omit the dependency on the initial condition $x$ of the process $X^x_\cdot$.


The above formulation of FBSDEs is intrinsically linked to the following stochastic optimal control problem:

\begin{align}\label{eq:min}
& \underset{y,\; Z=(Z_t)_{t\in [0,T]}}{\text{minimise}} \; \EE\left[ \left| \vartheta(X_T) - Y^{y,Z}_T\right|^2\right]\\
& \text{subject to } \begin{cases}
X _ { t }  &\hspace{-0.2cm} =   x + \int _ { 0 } ^ { t } b \left( s, X _ { s } \right) \mathrm d s + \int _ { 0 } ^ { t } a \left( s, X _ { s } \right)^\top \mathrm d W^{\QQ} _ { s } ,\\
Y^{y,Z}_{t}   
  &   \hspace{-0.2cm} = y  -\int_{0}^{t} h \big(s, X_{s}, Y^{y,Z}_{s}, Z_{s}\big) \mathrm d s +\int_{0}^{t} Z_{s}^\top \, \mathrm d W^{\QQ}_{s}, \quad t \in[0, T].
\end{cases}\label{eq:dinXY}
\end{align}

In particular, a solution $(Y,Z)$ to \eqref{eq:backward} is a  minimiser of the problem \eqref{eq:min}. A discretized version of the optimal control problem \eqref{eq:min}--\eqref{eq:dinXY} is the basis of the Deep BSDE Solver. 

Given $N\in \mathbb N$, consider  $0=t_0< t_1<\ldots <t_N=T$. For simplicity, let us take a uniform mesh with step $\Delta t$ such that $t_n = n\Delta t$, $n=0,\ldots,N$, and denote $\Delta W_{n} = W^{\QQ}_{t_{n+1}} - W^{\QQ}_{t_n}$. By an Euler-Maruyama approximation of \eqref{eq:min}--\eqref{eq:dinXY},  one has
\begin{align}
{\widetilde X}_{{n+1}}    & =  {\widetilde X}_{n}  + b(t_n,{\widetilde X}_{n}) \Delta t + a(t_n,{\widetilde X}_{n})  \Delta W_{n}, && \widetilde X_{0} = x, \label{eq:Eforward}\\
{\widetilde Y}^{y,\widetilde Z}_{{n+1}} &=  {\widetilde Y}^{y,\widetilde Z}_{n} - h(t_n, {\widetilde X}_{n}, {\widetilde Y}^{y,\widetilde Z}_{n}, \widetilde{Z}_{n}) \Delta t + \widetilde{Z}_{n}^\top\Delta W_{n}, && {\widetilde Y}^{y,\widetilde Z}_0 = y.\label{eq:Ebackward}
\end{align}

 The core idea of the Deep BSDE Solver is to approximate, at each time step $n$,  the control process $\widetilde Z_n$ in \eqref{eq:Ebackward} by using an artificial neural network (ANN). 
More specifically, in the Markovian setting, $Z_t$ is a measurable function of $X_t$, which we approximate by an ANN ansatz to carry out the optimisation above
 over this parametrised form.
 To this end, we introduce next a formalism for the description of neural networks.

\subsubsection*{ANN approximation}\label{sec:ANN}
We consider artificial neural networks with $\mathcal{L}+1\in\mathbb{N}\setminus \{1,2\}$ layers. Each layer consists of $\nu_\ell$ \textit{nodes} (also called \textit{neurons}), for $\ell=0,\ldots,\mathcal{L}$. The $0$-th layer represents the \textit{input layer}, while the $\mathcal{L}$-th 
layer is called the \textit{output layer}. The remaining $\mathcal{L}-1$ layers are \textit{hidden layers}. For simplicity, we set $\nu_\ell=\nu$, $\ell=1,\ldots,\mathcal{L}-1$. The input and output dimensions are both $d$ in our case.

A feedforward neural network is a function $\varphi^{\varrho}:\mathbb{R}^d\mapsto \mathbb{R}^{d}$, 
defined via the composition
\begin{align*}
x \in \mathbb{R}^{d} \longmapsto \mathcal{A}_{\mathcal{L}} \circ \varrho \circ \mathcal{A}_{\mathcal{L}-1} \circ \ldots \circ \varrho \circ \mathcal{A}_{1}(x) \in \mathbb{R}^{d},
\end{align*}  
where all $\mathcal{A}_\ell$, $\ell=1,\ldots,\mathcal{L}$, are affine transformations 
\begin{align*}
\mathcal{A}_1:\mathbb{R}^d \mapsto\mathbb{R}^\nu, \qquad
\mathcal{A}_\ell:\mathbb{R}^\nu \mapsto\mathbb{R}^\nu,  \quad  \ell=2,\ldots,\mathcal{L}-1,\qquad
\mathcal{A}_\mathcal{L}:\mathbb{R}^\nu\mapsto\mathbb{R}^{d},
\end{align*}
 of the form
$\mathcal{A}_\ell(x):=\mathcal{W}_\ell x + \beta_\ell$,  $\ell=1,\ldots,\mathcal{L}$,
where $\mathcal{W}_\ell$ and $\beta_\ell$ are matrices and vectors of suitable size called, respectively,  weights and biases. The function $\varrho$, called \textit{activation function} is a univariate function $\varrho:\mathbb{R}\mapsto\mathbb{R}$ that is applied component-wise to vectors. With an abuse of notation, we denote
$\varrho(x_1,\ldots,x_\nu)=\left(\varrho(x_1),\ldots,\varrho(x_\nu)\right).$
The elements of $\mathcal{W}_\ell$ and $ \beta_\ell$ are the parameters of the neural network. We can regroup all parameters in a vector $\rho\in\mathbb{R}^R$ where $R=\sum_{\ell=0}^\mathcal{L}\nu_\ell(1+\nu_\ell)$.

\begin{figure}
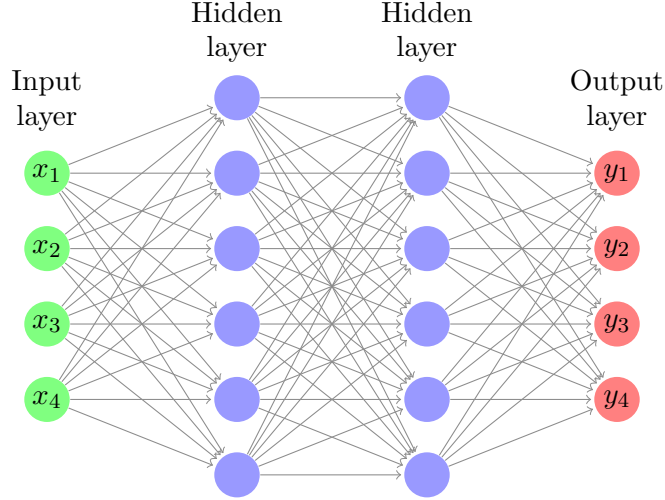

\centering
\vspace{1cm}
\begin{neuralnetwork}[height=4]
		\newcommand{\nodetextclear}[2]{}
		\newcommand{\nodetextx}[2]{$x_#2$}
		\newcommand{\nodetexty}[2]{$y_#2$}
		\inputlayer[count=4, bias=false, title=Input \\layer, text=\nodetextx]
		\hiddenlayer[count=6, bias=false, title={\vspace{-1.5cm}}\\Hidden layer, text=\nodetextclear] \linklayers
		\hiddenlayer[count=6, bias=false, title={\vspace{-1.5cm}}\\Hidden layer, text=\nodetextclear] \linklayers
		\outputlayer[count=4, title=Output layer, text=\nodetexty] \linklayers
\end{neuralnetwork}
\caption{Schematic representation of a feedforward neural network with two hidden layers, i.e.\ $\mathcal{L}=3$, input and output dimension $d=4$, and
$\nu=d+2=6$ nodes.}
\end{figure}

As indicated earlier, we use ANNs to approximate the control process $Z_t$. More specifically, let $R\in \NN$ as before and let $\xi\in \RR$, $\rho\equiv(\rho_1\ldots, \rho_R)\in \RR^{R}$  be $R+1$ parameters. We introduce a family of neural networks $\varphi^{\rho}_n :\RR^d \to \RR^d$, $n\in\{0,\ldots, N\}$ parametrized by  $\rho$ and indexed by time.  We denote   ${\mathcal Z}^{\rho}_{n} = \varphi^{\rho}_n(\widetilde X_{n})$ and consider the following parametrized version of \eqref{eq:Ebackward}
\begin{align}
{\mathcal Y}^{\xi,\rho}_{{n+1}} &=  {\mathcal Y}^{\xi,\rho}_{n} - h(t_n, X_{n},{\mathcal Y}^{\xi,\rho}_{n},{\mathcal Z}^{\rho}_{n}) \Delta t + ({\mathcal Z}^{\rho}_{n})^\top\Delta W_{n}, & {\mathcal Y}^{\xi,\rho}_0 = \xi,\label{eq:Ebackwardrho}
\end{align}
meaning that, at each time step, we use a distinct neural network to approximate the control process. 
The Deep BSDE Solver by \cite{ehanjen17}
 considers the following stochastic optimization problem 
\begin{align}\label{eq:minrho}
& \underset{\xi\in \RR,\; \rho \in \RR^{R}}{\text{minimise}}\; \;\EE\left[ \left(\vartheta(\widetilde X_N) - {\mathcal Y}^{\xi,\rho}_N\right)^2\right]\quad \text{subject to \eqref{eq:Eforward}--\eqref{eq:Ebackwardrho}}.
\end{align}
Observe that, in practice,  one simulates $L\in \NN$ Monte Carlo paths $(\widetilde X^{(\ell)}_n, {\mathcal Y}^{\xi,\rho,(\ell)}_n)_{n=0\ldots N}$ for $\ell=1,\dots,L$, using \eqref{eq:Eforward}--\eqref{eq:Ebackwardrho} with $N$ i.i.d.\ Gaussian random variables  $(\Delta W_n)_{n=0,\ldots,N-1}$ with  mean $0$ and variance $\Delta t$. Replacing the expected cost
functional by the empirical mean, \eqref{eq:minrho} becomes
\begin{align}\label{eq:minrhoell}
& \underset{\xi\in \RR,\; \rho \in \RR^{R}}{\text{minimise}} \;\; \frac{1}{L} \sum^L_{\ell=1}  \left(\vartheta(\widetilde X^{(\ell)}_N) - {\mathcal Y}^{\xi,\rho,(\ell)}_N\right)^2 \quad \text{subject to \eqref{eq:Eforward}--\eqref{eq:Ebackwardrho}}.
\end{align}
This minimization typically involves a huge number of parameters and it is performed by a  stochastic gradient descent-type algorithm (SGD), leading to random approximations. For further details on this point we refer the reader to Section 2.6 in \cite{ehanjen17}.  
We will denote by $\mathcal I$ the maximum number of SGD iterations. 
To improve the performance and stability of the ANN approximation, a \emph{batch normalization} is also considered, see \cite{IoffeSzegedy}.  

The accuracy of the solution is determined by the number of timesteps, number of samples, the chosen network architecture, and the quality of the optimiser found by the chosen optimisation routine. Our practical experience shows that quantifying and controlling the errors resulting from the latter two contributions is particularly difficult.
Therefore, 
certain \emph{a posteriori} error bounds as found in \cite{bender2013posteriori} for decoupled FBSDEs, \cite{HanLon18} for partially coupled FBSDEs, and
in \cite{reisinger2020posteriori} for fully coupled BSDEs are particularly valuable.
Specifically, in \cite[Theorem 1']{HanLon18} the authors show that under suitable assumptions on the coefficients of the FBSDE \eqref{eq:forward}--\eqref{eq:backward}, namely, in the decoupled case (see their Assumption 3, 2.), the uniform Lipschitz continuity in space, uniform 1/2-H\"older continuity in time of $b, a, h$ and the Lipschitz continuity of $\vartheta$) one has,  for $\Delta t$ sufficiently small,
\begin{equation}\label{eq:error}
\sup_{t\in [0,T]} \EE | Y_t -  Y^{\xi,\rho}_t |^2 + \int^T_0 \EE|  Z_t -  Z^{\rho}_t |^2 \, \mathrm d t   \leq   C\Big(\Delta t +  \EE\Big[ \left(\vartheta(\widetilde X_N) - {\mathcal Y}^{\xi,\rho}_N\right)^2\Big]\Big),
\end{equation}
where $C$ is a constant  independent of $\Delta t$ and $d$ possibly depending on the starting point $x$ of the forward process and,  given $({\mathcal Y}^{\xi,\rho}_{n}, {\mathcal Z}^\rho_n)_{n=0,\ldots, N}$ from \eqref{eq:Ebackwardrho}, $Y^{\xi,\rho}_t = {\mathcal Y}^{\xi,\rho}_{n}$ and $ Z^{\rho}_t = {\mathcal Z}^{\rho}_{n}$ for $t\in [t_n, t_{n+1})$.

In \cite[Theorem 2']{HanLon18}, \emph{a priori} estimates on the term $\EE[ (\vartheta(\widetilde X_N) - {\mathcal Y}^{\xi,\rho}_N)^2]$ appearing in the right hand side of \eqref{eq:error} are also provided. However, the obtained bounds depend on the (unknown) approximation capacity of the considered ANN. 

{
In addition to the combined error bound on the $Y$ and $Z$ approximations in \eqref{eq:error}, we can bound the error for $Y$ in terms of the error for $Z$, as shown in Appendix \ref{app:boundYZ}:
\begin{equation}\label{eq:errorYZ}
\sup_{t\in [0,T]} \EE | Y_t -  Y^{\xi,\rho}_t |^2 \le C
\left(
 \EE | Y_0 - \xi |^2 +
 \int^T_0 \EE|  Z_t -  Z^{\rho}_t |^2 \, \mathrm d t
 \right),
\end{equation}
for a constant $C$ that only depends on the model parameters (but not $\xi$ or $\rho$).
The controls of the Deep BSDE Solver directly influence the terms on the right-hand side by choice of $\xi$ and $\rho$.
The numerical tests in Section \ref{sec:numerics} indicate that in our applications $\xi$ can typically be determined more easily and accurately
than $\rho$, and that the errors in $Y$ and $Z$ are of similar magnitude.
}

\begin{remark} Let us comment on our use of the the Deep {BSDE} Solver.
\begin{itemize}
\item {The new and distinctive feature of our approach consists in employing the deep solver to perform scenario simulations, i.e.\ simulations of the evolution of the mark-to-market of a portfolio of claims. This shows, for the first time, that the deep solver can be successfully used to solve risk-management problems such as the calculation of xVAs and risk measures (e.g. Value-at-Risk, Expected Shortfall).}
\item {Our use of the Deep BSDE Solver provides an alternative to nested Monte Carlo simulations and  their competitors such as regression Monte Carlo (see \cite{LonSch01}). Indeed, for the calculation of xVAs we are not only interested in the initial value of the BSDE solution {(which is the optimal $\xi$ given as output by the Deep BSDE Solver)}, but we need the ability to simulate the evolution of the conditional expectations representing the mark-to-market of the portfolio.}
\item {We also observe that our algorithm is interesting in comparison to, e.g., the commonly used regression approaches because we do not only solve for the evolution of the value function {(i.e. the portfolio value)}, but we also obtain the control {(i.e., the hedging strategy)}, which is linked to the sensitivies of the value function. This is of paramount importance in order to calculate many risk measures such as initial margin according to the ISDA Simm methodology. This feature means that our use of the Deep BSDE Solver is useful also in a low dimensional setting, where traditional numerical techniques suffering from the curse of dimension are still viable.}
\end{itemize}

\end{remark}

\subsection{The Deep xVA Solver for non-recursive valuation adjustments}
\label{subsec:nonrec}

In our setting, the Deep BSDE Solver is first employed in the approximation of the clean values of the portfolio, i.e., the processes $\widehat V^m_t$ for $m=1,\ldots, M$, which are the solutions of \eqref{cleanBSDEeu} with underlying forward dynamics given by $S$ in \eqref{eq:asset_prox}. More precisely, in the notation of the previous section, we take 
$$
X_t = S_t \quad\text{and}\quad Y_t = \widehat V^m_t \quad \text{ for } m=1, \ldots, M.
$$
For simplicity, let us assume $T_m=T, \;\forall m=1,\ldots, M$.
We now describe the algorithm for computing CVA and DVA given by formulas \eqref{eq:defCVA} and \eqref{eq:defDVA}, respectively.  A unifying formula for CVA and DVA can be written as
\begin{equation}\label{eq:VAgeneral}
\Excond{\QQ}{ \int^T_t \Phi_u(\widehat V_u) \, \mathrm d u }{\cF_t},
\end{equation}
where 
\begin{itemize}
\item $ \Phi_u(v) = (1-R^C)\frac{B^{\tilde{r}}_t}{B^{\tilde{r}}_u}\left(v -C(v) \right)^-\lambda^{C,\QQ}_u $ for CVA;
\item $\Phi_u(v) = (1-R^B)\frac{B^{\tilde{r}}_t}{B^{\tilde{r}}_u}\left(v- C(v) \right)^+\lambda^{B,\QQ}_u$  for DVA.
\end{itemize}
Here, $\Phi_u(v)$ indicates that $\Phi$ is a random field. One can easily observe that, thanks to the boundedness of the processes $\tilde r$ and $\lambda^j$, $j\in \{B,C\}$,  $\Phi_u(v)$ is uniformly Lipschitz continuous in $v$. We denote by $L_\Phi$ its Lipschitz constant.

Given a time discretization (uniform, for simplicity) with time step $\Delta t$, the integral in \eqref{eq:VAgeneral} can be approximated by a quadrature rule, i.e., 
taking $t=t_0=0$,
$$
\int^T_{0} \Phi_u(\widehat V_u) \, \mathrm d u \approx \sum^N_{n=0} \eta_n \Phi_{t_n}(\widehat V_{t_n}).
$$
For instance, 
one may consider the rectangle rule, i.e. $\eta_N=0, \eta_n=\Delta t \; n=0,\ldots, N-1$,
\begin{equation}\label{eq:trapez}
\int^T_0 \Phi_u(\widehat V_t) \, \mathrm d t \approx  \sum^{N-1}_{n=0}   \Phi_{t_n}(\widehat V_{t_n}) \Delta t.
\end{equation}
Denoting for any $m=1,\ldots, M$ by $\big(\widehat{\cV}^{m,{\xi}^*_m,{\rho}^*_m, (p)}_n\big)_{n = 0,\ldots, N, p = 1,\ldots, P}$   the approximation of {$P$ paths of}  the process $(\widehat V^m_{t_n})_{n=0,\ldots, N}$ obtained by means of the  parameters $({\xi}^*_m, {\rho}^*_m)$ resulting from the Deep BSDE Solver optimization \eqref{eq:minrhoell} and 
$$
\widehat{\cV}^{\,*,(p)}_n  := \sum^M_{m=1} \widehat{\cV}^{\,m,{\xi}^*_m,{\rho}^*_m, (p)}_n, \qquad n=0,\ldots N,
$$
the adjustment is then approximated by the following formula:
$$
 \frac{1}{P} \sum^P_{p=1} \sum^N_{n=0} \eta_n \Phi_{t_n}(\widehat{\cV}^{\,*,(p)}_n ).
$$
{Here, $P$  denotes the number of Monte Carlo paths used for estimating the outer expectation in \eqref{eq:VAgeneral} which are typically different from the $L$ paths generated for  training the NN.}
Algorithms \ref{algo:solver} and \ref{algo:solverForNonRecursive} summarize the main steps of the method.
In what follows we will also denote by $\widehat{V}^{m,{\xi}^*_m,{\rho}^*_m}$ the piecewise constant interpolation of $\widehat{\cV}^{m,{\xi}^*_m,{\rho}^*_m}$.

\SetKwFunction{Fn}{{\bf Deep BSDE Solver}}

\begin{algorithm}[h]
\SetAlgoLined
\SetKwFunction{FMain}{Main}
 \SetKwProg{Fn}{Deep BSDE Solver}{:}{}
 Set parameters: $N, L, { B}$.   \Comment{\textit{$N$ time steps, $L$ paths for inner  Monte Carlo loop, { $B$ batch size}}}\\
 Fix architecture of ANN. \Comment{\textit{intrinsically defines the number of parameters $R$}}\\
 
\Fn{($N$,$L$,{ $B$})}{\ \\
  Simulate $L$ paths $(\widetilde S^{(\ell)}_n)_{n=0,\ldots, N}$, $\ell=1,\ldots,L$ of the forward dynamics. \\
  Define the neural networks $(\varphi^{\rho}_n)_{n=1,\ldots, N}$. \\
  \For{$m=1,\ldots, M$}{ Minimize over $\xi$ and $\rho$
$$
\frac{1}{L} \sum^L_{\ell=1}\left( g_m(\widetilde S^{(\ell)}_N) - \widehat{\cV}^{m,\xi,\rho,(\ell)}_N\right)^2,
$$ 
\qquad\qquad subject to 
\begin{equation}\label{alg:bsde}
\left\{
\begin{split}
\widehat{\cV}^{m,\xi,\rho,(\ell)}_{{n+1}} & =  \widehat{\cV}^{m,\xi,\rho,(\ell)}_{n} + r_{t_n}  \widehat{\cV}^{m,\xi,\rho,(\ell)}_{n} \Delta t + (\widehat{\cZ}^{m, \rho,(\ell)}_{n})^\top\Delta W^{(\ell)}_{n},\\
\widehat{\cV}^{m,\xi,\rho,(\ell)}_0 &  = \xi, \\
 \widehat{\cZ}^{\rho, (\ell)}_{n}  & = \varphi^\rho_n(\widetilde S^{(\ell)}_n).
 \end{split}
 \right.
\end{equation}
Save the optimizer $(\xi^*_m,\rho^*_m) $.
}
}
{\bf end}\\
\caption{Deep algorithm for exposure simulation}\label{algo:solver}
\end{algorithm}
\begin{algorithm}[h]
\SetAlgoLined
\SetKwFunction{FMain}{Main}
 \SetKwProg{Fn}{Deep BSDE Solver}{:}{}
Apply Algorithm \ref{algo:solver} \\

Set parameters: $P$.   \Comment{\textit{$P$ paths for the outer Monte Carlo loop}}\\

Simulate, for \hspace{-0.05cm}$m=1\ldots M$,  $\big(\widehat{\cV}^{\,m,{\xi}^*_m,{\rho}^*_m, (p)}_n\big)_{n=0 \ldots N, p=1\ldots P}$ by means of  \eqref{alg:bsde} with \hspace{-0.1cm}$(\xi,\rho)=(\xi^*_m, \rho^*_m)$. \\
\Comment{\textit{approximation of the clean values }\\}
Define $\widehat {\cV}^{\,*,(p)}_n := \sum^M_{m=1} \widehat{\cV}^{\,m,{\xi}^*_m,{\rho}^*_m, (p)}_n$ for  $n=0\ldots N$, $p=1\ldots P$. \\
\Comment{\textit{approximation of the clean portfolio value}\\}
Compute the adjustment as 
$$
\frac{1}{P} \sum^P_{i=1} \left( \sum^N_{n=0} \eta_n \Phi_{t_n}(\widehat{\cV}^{\, *,(p)}_n) \right). 
$$
\caption{Deep  xVA Solver for non-recursive valuation adjustments}\label{algo:solverForNonRecursive}
\end{algorithm}

Under reasonable assumptions, we can derive the following \emph{a posteriori} bounds for the error associated with this approximation of the valuation adjustments in $[0,T]$, starting from \eqref{eq:error}. The derivation is given in Appendix \ref{app}. 
We note that these adjustments can also be obtained from the more general framework in Section \ref{subsec:rec},
however, we provide a simpler numerical procedure here and derive error estimates for these approximations by a more explicit computation.

Let $\widehat V_t = \sum_{m=1}^M\widehat{V}^m_t$ with $\widehat{V}^m_t$ given by \eqref{cleanBSDEeu},
and $\widehat{\mathcal V}^{\,*}_n = \sum^M_{m=1} \widehat{\mathcal V}^{m,\xi^*_m,\rho^*_m}_n$ ($n=0,\ldots, N)$ its approximation from the Deep BSDE Solver.
Consider the running assumptions of this paper together with uniform H\"older continuity in $t$ of 
$b$ and $\sigma$, and assume estimate \eqref{eq:error} for equation \eqref{cleanBSDEeu}.\footnote{
This is a straightforward extension of \cite[Theorem 1']{HanLon18} in the case of deterministic $r_s$ in \eqref{cleanBSDEeu}
which is H{\"o}lder-1/2 in $s$, by replacing their assumption on the uniform H{\"o}lder-1/2 continuity of $f$ by  $|f(t,v)-f(s,v)|\leq C |t-s|^{1/2} |v|$ for all $0\le s\le t\le T$
and all $v$.
 For stochastic rates, a more substantial extension to their analysis is needed for a direct application of Euler-Maruyama, due to the non-Lipschitz term $r_t \widehat V_t$ in \eqref{cleanBSDEeu} and accounting for the discretisation of the rates process. However, the simple transformation \eqref{trans} from the appendix can eliminate this drift.
We hence directly assume \eqref{eq:errorapp} for this analysis.}

Moreover, consider the specific forms of $\Phi$ above, assuming
$\mathbb{E}[(\lambda^{j,\QQ}_s-\lambda^{j,\QQ}_t)^2] \le C (t-s)$ for $0\le s\le t\le T$ and $j\in\{B,C\}$.\footnote{This is immediate for deterministic H{\"o}lder-1/2 functions and a standard property of It{\^o} diffusions with Lipschitz coefficients
(see \cite[Lemma 2.4, (2.10)]{Zhang04}), but also holds, e.g., for the Cox-Ingersoll-Ross process (as follows e.g.\ from \cite[Corollary 2.14]{hutzenthaler2014strong}).}

Then, for $\Phi$ as above, there exists a constant $C\geq 0$ depending only on the model inputs 
and the constants coming from \eqref{eq:error} (in particular not on $\Delta t$ and the ANN parameters),  such that
\begin{align}
\label{eqn:nonrec_error}
\bigg| \EE\Big[ \int^T_0  \Phi_t(\widehat V_t) \, \mathrm d t\Big] -  \EE\Big[ \sum^{N-1}_{n=0} \Delta t \Phi_{t_n}(\widehat{\cV}^{\, *}_{n}) \Big] \bigg| 
\leq C  \Big( \Delta t + \sum^M_{m=1} \EE\Big[ | g_m(S_{T}) -\widehat{\cV}^{\, m,\xi^*_m,\rho^*_m}_{N}|^2\Big] \Big)^{1/2}.
\end{align}

MC standard errors for the second expectation in \eqref{eqn:nonrec_error} should be added to obtain a complete bound.

\subsection{The Deep  xVA Solver for recursive valuation adjustments}
\label{subsec:rec}

The procedure of the previous section is sufficient to perform the estimation of CVA and DVA according to \eqref{eq:defCVA} and \eqref{eq:defDVA} at time zero by means of a standard Monte Carlo estimator, given the pathwise solutions of the BSDEs for clean values. 
Typically, however, the bank needs to also compute risk measures on the CVA, such as Value--at--Risk. Moreover, if we consider the xVA BSDE \eqref{eq:XVApreDef}, we observe that FVA terms introduce a recursive structure through the driver, so that a time $t$ estimate of the process $\overline{\rm XVA}$ requires the use of a numerical solver for a BSDE. Finally, let us observe that the bank is not only interested in computing the xVA at time $t$, also hedging the market risk of xVA is important, meaning that one also needs sensitivities of valuation adjustments with respect to the driving risk factors.

All above considerations motivate us to propose a two-step procedure, where we first employ the Deep BSDE Solver to estimate the clean values $\widehat{V}^m$, $m=1,\ldots,M$, according to Algorithm \ref{algo:solver} and then, using the simulated paths of the $M$ clean BSDEs obtained from the first step, we apply again the Deep BSDE Solver to numerically solve the xVA BSDE \eqref{eq:XVApreDef}. The procedure is outlined in Algorithm \ref{algo:solverForXva}.

\begin{algorithm}[h]
\SetAlgoLined
\SetKwFunction{FMain}{Main}
 \SetKwProg{Fn}{Deep XVA-BSDE solver}{:}{}
Apply Algorithm \ref{algo:solver}.

 Set parameters:  $P$.  \Comment{\textit{$P$ paths for outer Monte Carlo loop}}\\
 Fix architecture of ANN. \\
 \Comment{\textit{intrinsically defines the number of parameters $\bar R$ (in general $\bar R\neq R$)}}\\
\Fn{($N$,$P$)}{\ \\
  Simulate $P$ paths $(\cV^{(p)}_n)_{n=0,\ldots, N}$, $p=1,\ldots,P$, of the portfolio value. \\
  Define the neural networks $(\psi^{\zeta}_n)_{n=1,\ldots, N}$. \\
  \vspace{0.3cm}
  Minimize over $\gamma$ and $\zeta$
$$
\frac{1}{P} \sum^P_{p=1}\left( \overline{\cX}^{\gamma,\zeta,(p)}_N\right)^2,
$$ 
\qquad\qquad subject to 
\begin{equation}\label{alg:bsdeNumXVA}
\left\{
\begin{split}
\overline{\cX}^{\gamma,\zeta,(p)}_{{n+1}} & = \overline{\cX}^{\gamma,\zeta,(p)}_{n} - \bar f(t_n, \widehat{\cV}^{(p)}_{n}, \overline{\cX}^{\gamma,\zeta,(p)}_{n}) \Delta t + (\overline{\cZ}^{ \zeta,(p)}_{n})^\top\Delta W^{(p)}_{n},\\
\overline{\cX}^{\gamma,\zeta,(p)}_0 &  = \gamma, \\
\overline{ \cZ}^{\zeta,(p)}_{n}  & = \psi^\zeta_n(\widehat{\cV}^{(p)}_n).
 \end{split}
 \right.
\end{equation}
}
{\bf end}\\
\caption{Deep  xVA Solver }\label{algo:solverForXva}
\end{algorithm}

Similar to Section \ref{subsec:nonrec}, we can quantify the error of the Deep xVA Solver in the recursive case {\it a posteriori}.
Let $(\overline{\rm XVA}_t, \overline Z_t)$ be the solution of \eqref{eq:XVApreDef}, $(\widetilde {\rm XVA}^{\gamma,\zeta}_t, \widetilde Z^{\zeta}_t)$ the corresponding approximation from the Deep BSDE Solver with parameters 
$\gamma, \zeta$, with $\widehat{V}$ in \eqref{eq:XVApreDef} replaced by $\widehat  V^{\xi, \rho}$ given by the solver with parameters $\xi, \rho$.
We note that the result of \cite{HanLon18} can be extended to multi-dimensional BSDEs (see the comment at the start of Section 2 there), or that our system is a special case of the fully-coupled McKean--Vlasov FBSDEs analysed in \cite{reisinger2020posteriori} (where the monotonicity condition H.1.(1) imposed there is not needed here in the weakly coupled case).

Take the running assumptions of this paper. 
Moreover, let for simplicity all rates and intensity processes be bounded, uniformly 1/2-H\"older continuous deterministic functions of time
and the functions $\mu$, $\sigma$, $\bar f$ be uniformly 1/2-H\"older continuous in time.
Then there exists a constant $K\geq 0$ depending only on the model inputs (in particular not on $\Delta t$ and the ANN parameters) such that
\begin{align*}
&\sup_{t\in [0,T]} \EE \Big[\,\Big|\overline{\rm XVA}_t  - \widetilde {\rm XVA}^{\gamma,\zeta}_t\Big|^2\Big]  + \EE\left[ \int^T_0 | \overline Z_t - \widetilde Z^{\zeta}_t|^2 \, \mathrm d t\right] \\
& \leq K \left( \Delta t + \sum^M_{m=1} \EE\Big[ | g_m(S_{T}) -\widehat{\cV}^{\, m,\xi_m,\rho_m}_{N}|^2\Big]\; + \EE\Big[ |\widetilde {\rm XVA}^{\gamma,\zeta}_T|^2\Big]\right) .
\end{align*}

It should be possible to derive similar results for bounded or even unbounded stochastic rates, but care would have to be taken with the discretisation in the case of non-Lipschitz coefficients, such as the CIR model.

{
\subsection{Calculation of risk measures}\label{sec:riskMeasures} An important benefit of the deep xVA solver is given by the ability to compute risk measures as a by-product without additional numerical burden and to do so for any time horizon within the simulation time grid. More specifically, let $\mathcal{P}$ denote a process of interest, which could represent either the clean value $\widehat{\mathcal{V}}^{*,(p)}$ as estimated via Algorithm \ref{algo:solver} or the xVA correction $\overline{\mathcal{X}}^{*,(p)}$ as produced by Algorithm \ref{algo:solverForXva}. Given time points $t_n$, $n=1,\ldots,N$, we can define the loss {process}
\begin{align*}
L_{t_n}:=-\left(\mathcal{P}_{t_n}-\mathcal{P}_{t_0}\right).
\end{align*}
The above defined discrete time stochastic process $L$ can then be used to compute risk measures at each point in time over the simulation grid. To provide examples, we can compute e.g. 
\begin{itemize}
\item Value at Risk:
\begin{align*}
\text{VaR}_\alpha(\mathcal P_{t_n}):&=\inf\left\{\left. l \in\mathbb{R}\right| \mathbb{Q}\left(\left.L_{t_n}>l\right|\mathcal{F}_{t_0}\right)\leq 1-\alpha\right\},
\end{align*}
\item Expected Shortfall: 
\begin{align*}
ES_\alpha(\mathcal P_{{t_n}})&:=\mathbb{E}\left[\left.L_{{t_n}}\right|L_{{t_n}}\geq \text{VaR}_\alpha(L_{{t_n}}),\mathcal{F}_{t_0}\right],\\
\end{align*}
\end{itemize}
both on the clean value and, more importantly, on the xVAs. Notice that the computation of risk measures, e.g.\ on the CVA, does not require the use of nested simulations. We simply simulate the trajectories of the BSDE satisfied by the value adjustment and evaluate the risk measure over the simulated paths. This is demonstrated in Section \ref{subSec:basketCall}, where we compute the Value at Risk on the CVA of a $100$-dimensional basket option.}

\subsection{Pathwise simulation of sensitivities}\label{sec:sens}

 One interesting feature of our approach to xVA computations is that we can easily estimate several sensitivities (i.e., partial derivatives) of pricing functions. Let us recall that, in the present Markovian setting, the control $Z$ associated with a FBSDE  of the general  form \eqref{eq:forward}--\eqref{eq:backward} satisfies
\begin{align}\label{eq:hessY}
Z_t=\frac{\partial Y}{\partial X}(t,X_t)a(t,X_t),
\end{align}
so that we can easily reconstruct the gradient of the pricing function with respect to all risk factors simply by multiplying each (vector-valued) neural network by the inverse (assuming it exists) of the matrix $a(t,X_t)$.
This becomes particularly interesting in view of Algorithms \ref{algo:solver} and \ref{algo:solverForXva},
where we can obtain hedge ratios both for the clean value and for the valuation adjustments
without further computations.

Obtaining second order sensitivities, which may also be important for hedging purposes,  is also feasible in our setting, because feedforward neural networks are compositions of simple functions and computation of gradients of neural network functions has become standard in that community.
Using the notation of Section \ref{sec:ANN}, we can write
\begin{equation}
\label{eq:chainRule}
\frac{\partial{\mathcal Z}^\rho_n}{\partial X_n}=\frac{\partial \varphi^{\varrho}(X_n)}{\partial X_n},
\end{equation}
with $ \varphi^{\varrho}(X_n) = \mathcal A_{\mathcal L} (\rho (  \mathcal A_{\mathcal L-1} \ldots \rho(\mathcal A_1(X_n))))$. 
Since $(\mathcal{A}_\ell)_{\ell=1,\ldots, \mathcal L}$ are affine functions, their Jacobians are given by the weight matrices, i.e.
\begin{align*}
J_{\mathcal{A}_\ell}(\cdot)=\mathcal{W}_\ell, \qquad \ell = 1,\ldots, \mathcal L.
\end{align*}
Moreover, one also has  
the Jacobian of $\rho$,
\begin{align*}
J_{\varrho}(\cdot) =\diag\left(\varrho^\prime (\cdot)\right),
\end{align*}
where, for $x\in \RR^\nu$ we denote $\varrho'(x) = (\varrho'(x_1),\ldots , \varrho'(x_\nu))$. 
In the present paper, we choose 
$\varrho(x)=\text{ReLU}(x)=\max\{x,0\}$ so that the first derivative   can be defined as
\begin{align*}
\varrho^\prime(x)=\operatorname{ReLU}^{\prime}(x)=\left\{\begin{array}{ll}{1} & {\text { if } x>0} \\ {0} & {\text {otherwise }}\end{array} \right\}  = \, \operatorname{sgn}(\operatorname{ReLU}(x)).
\end{align*}
Finally, we deduce that the following explicit differentiation formula holds:
\begin{align*}
\frac{\partial {\mathcal Z}^\rho_n}{\partial X_n} = \mathcal{W}_{\mathcal L}\;  \diag\left(\varrho^\prime (\mathcal{A}_{\mathcal L -1}(\ldots \mathcal A_1(X_n ))  )\right) \; \ldots\;  \diag\left(\varrho^\prime (\mathcal{A}_{1}(X_n))  \right)   \mathcal{W}_{1}. 
\end{align*}
Given the availability of the derivative of ${\mathcal Z}^\rho_n$ we can then obtain the Hessian of $Y$ {from \eqref{eq:hessY}}.

\section{Numerical results}\label{sec:numerics}

To test our algorithm, we start by studying two very simple examples with a similar computational structure as CVA and DVA, and for which we can easily provide reference solutions. We will then give a higher-dimensional example and illustrate further practically relevant features of the method, such as recursive xVA computations
and simulation of the collateral account. The codes for the proposed experiments are available at \url{https://github.com/AlessandroGnoatto/Deep-xVA-Solver}. \medskip

Let $S$  be the price of a single stock described by a  Black-Scholes dynamics,
$$
\mathrm d S_t = r S_t \, \mathrm d t + \sigma S_t \, \mathrm d W^{\QQ}_t,\qquad S_0=s_0,
$$ 
and $\widehat V$ a European-style contingent claim with value
$$
\widehat V_t = \EE^{\QQ}\left[ e^{-r (T-t)} g(S_T) | \cF_t\right].
$$
In particular, $\widehat V$ solves the following BSDE:
\begin{align}
\begin{cases}
-\mathrm d\widehat V_t =  -r \widehat V_t \, \mathrm d t   - \widehat Z_t \, \mathrm dW^{\QQ}_t, \\
\widehat V_T = g(S_T).
\end{cases}
\end{align}
 The discounted  expected positive and  negative 
 exposure of $\widehat V$ are defined, respectively,  by 
\begin{align}
\mathrm{DEPE}(s) & = \Excond{\QQ}{ e^{-r(s-t)} \left(\widehat V_s\right)^+}{\cF_t},\label{eq:defDEPE}\\
\mathrm{DENE}(s) & = - \Excond{\QQ}{ e^{-r(s-t)} \left(\widehat V_s\right)^-}{\cF_t}.\label{eq:defDENE}
\end{align}

In order to take into account the randomness of the algorithm (through the inner and outer Monte Carlo estimation and stochastic gradient descent), in the plots below we report with solid lines the average DEPE (in blue) and DENE (in red) obtained after 100 runs of the algorithm and the gray region represent the obtained standard deviation from the average value.

\subsection{A forward on $S$}\label{sec:fwdExample}
In this case, we consider 
$$
g(S_T) = S_T -K
$$
with $K=s_0$. The pathwise exposure $\widehat V$ at time $s\in [t, T]$ is given by 
\begin{align*}
\widehat V_s = \Excond{\QQ}{e^{-r(T-s)} (S_T-K)}{\cF_s} 
       = S_s  - K e^{-r(T-s)}.
\end{align*}

Substituting in \eqref{eq:defDEPE}, one has 
\begin{align}
\mathrm{DEPE}(s) 
\label{eq:exact_DEPEforward}
& = S_t \Phi(d_1) - K e^{-r(T-t)} \Phi(d_2), \\
\mathrm{DENE}(s) 
\label{eq:exact_DENEforward}
& = S_t \Phi(-d_1) - K e^{-r(T-t)}  \Phi(-d_2), 
\end{align}
where $\Phi(\cdot)$ denotes the standard normal cumulative distribution function and, as usual, 
$$
d_1= \frac{\ln\left(e^{r(t-s)} \, S_t/K\right) + \left(r +\sigma^2/2 \right) (s-t) }{\sigma\sqrt{s-t}}\quad\text{and}\quad d_2= d_1 -\sigma \sqrt{s-t}.
$$

\begin{table}[!hbtp]
\centering
\begin{tabular}{|c|c|c|c|c|c|c|}
\hline 
 $\sigma$ &  $K$ & $T$\\
\hline
 $0.25$  & $100$  & $1$\\
\hline
\end{tabular}
\caption{Parameters used in numerical experiments.}\label{table:param}
\label{tab:data}
\end{table}

We report in Figure \ref{fig:forward} the plot of the numerical results obtained by Algorithm \ref{algo:solverForNonRecursive} using the parameters in Table \ref{table:param} and $r=0$.  In particular, on the left we plot the simulated pathwise exposure, i.e.\ the paths  $t_n\rightarrow \widehat{\cV}^{*,(p)}_{n}$ for $p=1,\ldots, P$ {obtained by a single run of the algorithm},
while on the right we compare the approximated DEPE and DENE (solid lines) with the exact expected exposures given by \eqref{eq:exact_DEPEforward}--\eqref{eq:exact_DENEforward}  (dashed lines).  {The maximum difference between the approximated expected exposure and the exact one is $8.1$ bps for the DEPE and $12$ bps for the DENE (in both cases achieved at the terminal time $T$) with a maximal standard deviation of $0.3647$. }\\
{We asses the performance of the solver in the reconstruction of exposure trajectories computing the average terminal square error on the exposure, i.e.  $\frac1P \sum^P_{p=1} ( \widehat{\cV}^{*,(p)}_{N} - (S^{(p)}_T -K))^2$ which, for a chosen single run of the algorithm, is found to be $0.1883$ (comparable with the loss function given by the solver, which is  $0.1664$). {Moreover, in this special case, we can also compare $\widehat{\cZ}^{\rho^*}$ with the exact hedging strategy  $\widehat Z_t = \sigma S_t$. {We display in Figure \ref{fig:control_error} the exact and approximated hedging strategy, for which we obtain an  $L^2$-norm of the error equal to  $0.1759$. {The CPU time is $503$ s.} }}

\begin{figure}
\includegraphics[width=0.49\textwidth]{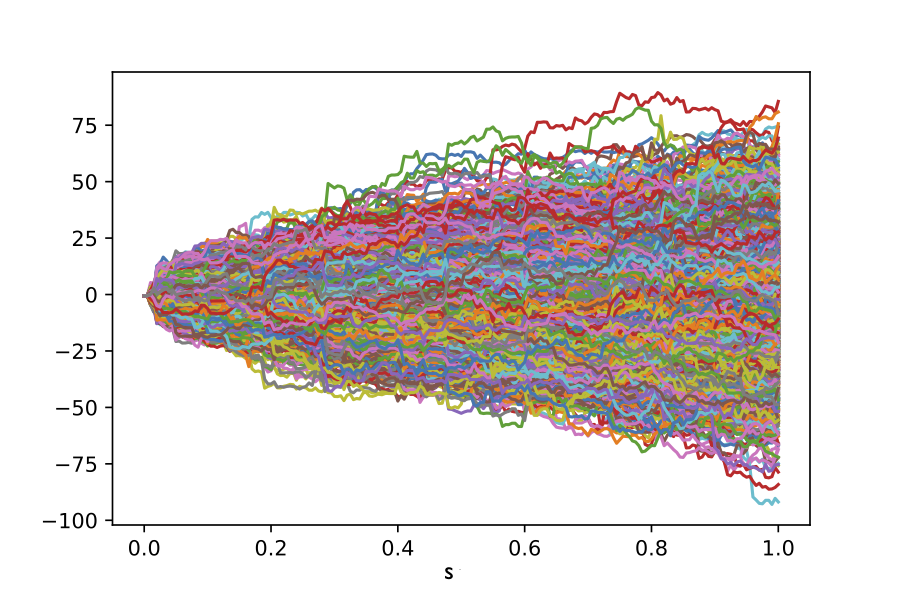}
\includegraphics[width=0.43\textwidth]{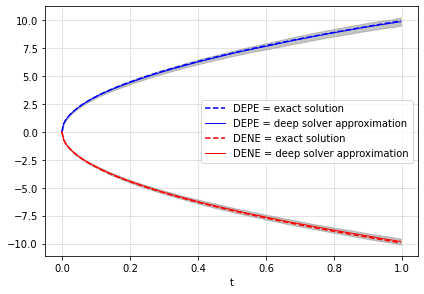}
 \caption{Forward contract: approximated exposure (left) and DEPE, DENE (right). Parameters used: outer MC paths $P = 2048$, inner MC paths $L=1024$, { batch size $B=64$}, 
 internal layers $\mathcal{L}-1 =  2$, {  nodes of each internal layer} $\nu = d+20 = 21$, $ \mathcal I$ = 4000, time steps $N = 200$.}
 \label{fig:forward}
\end{figure}

\begin{figure}
\includegraphics[width=0.45\textwidth]{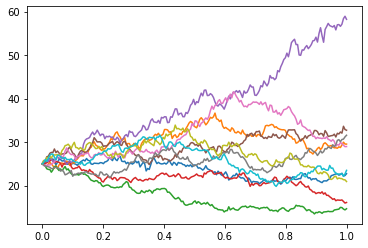}
\includegraphics[width=0.45\textwidth]{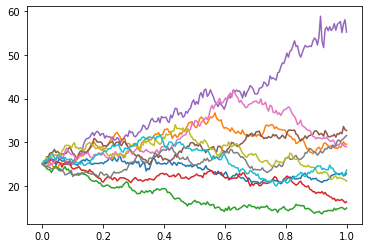}\\
 \caption{The  hedging strategy  for the forward contract on 10 simulated scenarios: exact  (left) and approximated  (right).} 
 \label{fig:control_error}
\end{figure}

\subsection{A European call option}\label{sec:call}
In this case we consider 
$$
g(S_T) = \left(S_T -K\right)^+,
$$
where we set $K=s_0$. The pathwise exposure $\widehat V$ at time $s\in [t, T]$ is given by the Black-Scholes formula
$$
\widehat V_s = \Excond{\QQ}{e^{-r(T-s)} \left(S_T - K\right)^+}{\cF_s} = S_s \Phi(d_1) - K e^{-r(T-s)} \Phi(d_2)>0.
$$
It follows immediately that  
$$
\mathrm{DEPE}(s) = \Excond{\QQ}{ e^{-r(s-t)} \widehat V_s }{\cF_t} = \widehat V_t,
$$
and
$$
\mathrm{DENE}(s) = 0.
$$

{The average terminal square error on the exposure computation, for a chosen run of the algorithm, is given by $0.7894$ (comparable with the loss function given by the solver  $0.7779$). {In this case, the exact hedging strategy  is  $\widehat Z_t =  \sigma \Phi(d_1) S_t$, from which we can compute  the  $L^2$-norm of the error with the approximated control  $\widehat{\cZ}^{\rho^*}$ which  is equal to  $0.1496$. We report in Figure \ref{fig:control_error_call} the exact and the approximated hedging strategy.}}
\\
The results obtained using   Algorithm \ref{algo:solverForNonRecursive} with  the parameters in Table \ref{table:param} and $r=0.01$ are reported in Figure \ref{fig:call} (left). The exact European  call price is $10.4036$, while the approximation of the positive and negative exposure obtained by the solver and reported in Figure \ref{fig:call} (left) take values, for $t\in [0,T]$, within the interval $[10.4072, 10.4963]$  and $[-0.1692, 0]$, respectively. The accuracy of the time zero option value for this architecture and simulation parameters is hence 0.36 bps, and that of DEPE and DENE in the worst case (over $s$) is 9.3 bps and 17 bps, respectively. {The CPU time is $543$ s.}

{
We also report in Figure \ref{fig:callVar} the approximation of the Value at Risk ($\text{VaR}_\alpha(\widehat V_t)$, with { $\alpha=0.05$}) computed as explained in Section 3.4.
Comparing with the exact values of the VaR (dashed line in Figure \ref{fig:callVar}), one can observe a good fit. {We point out that  for the VaR approximation we imposed the positivity of the value $\widehat V$, since without this condition precision was lost close to maturity}.}
}

\begin{figure}
\includegraphics[width=0.43\textwidth]{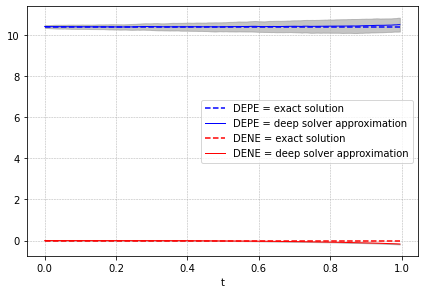}
\includegraphics[width=0.43\textwidth]{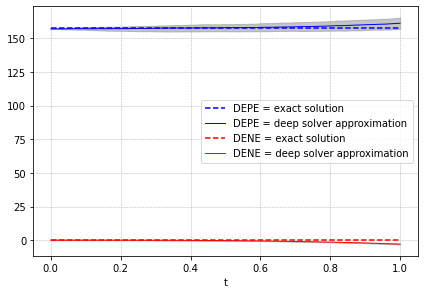}
 \caption{DEPE and DENE for a European call option (left) and a European basket option with 100 underlyings (right). Parameters used: outer MC paths $P = 2048$, inner MC paths $L=1024$, { batch size $B=64$}, internal layers $\mathcal{L}-1=  2$, { nodes of each internal layer} $\nu=d+20=21$ (left) and $\nu = d+10 = 110$ (right), iterations $\mathcal I$ = 4000 (left) and $\mathcal I$ = 10000 (right), time steps $N = 200$ (left) and $N=100$ (right).}\label{fig:call}
\end{figure}

\begin{figure}
\includegraphics[width=0.45\textwidth]{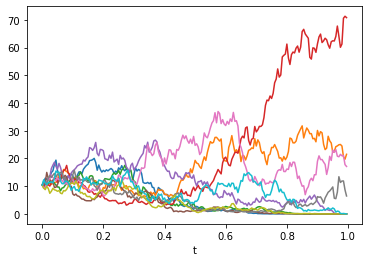}
\includegraphics[width=0.45\textwidth]{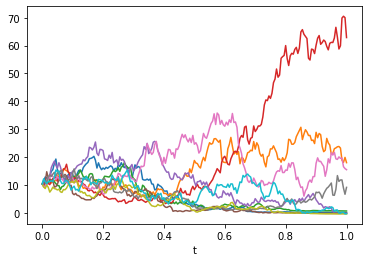}\\
\includegraphics[width=0.45\textwidth]{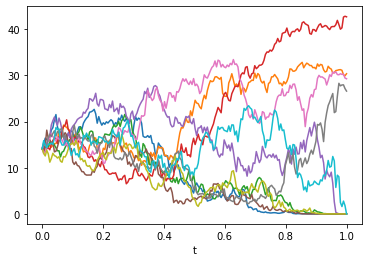}
\includegraphics[width=0.45\textwidth]{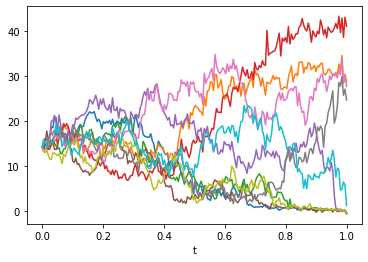}
 \caption{The {clean value (top row)} and hedging strategy   {(bottom row)} for the call option on 10 simulated scenarios: exact  (left) and approximated  (right).} 
 \label{fig:control_error_call}
\end{figure}

\begin{figure}
\includegraphics[width=0.43\textwidth]{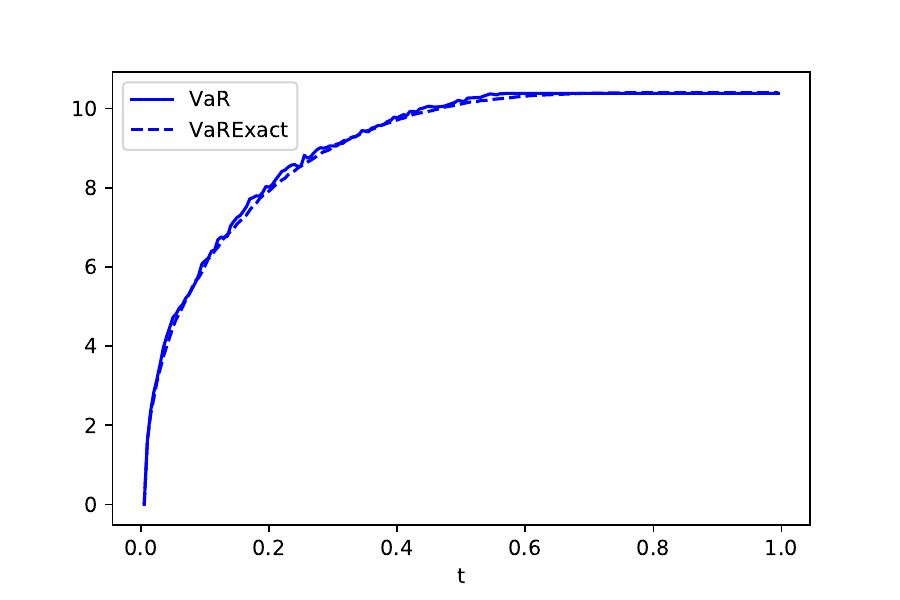}
 \caption{{Approximation of the VaR$(\widehat V_t)$ for $t\in [0,T]$. Parameters used: see the caption of Figure \ref{fig:call} (left)}.}\label{fig:callVar}
\end{figure}

\subsection{A basket call option}\label{subSec:basketCall}
Let us now consider the case of several underlying assets $(S^1, \ldots, S^d)$:
$$
\mathrm d S^i_t = r^i S^i_t \, \mathrm d t + \sigma^i S^i_t \, \mathrm d W^{\QQ,i}_t,\qquad S^i_0=s^i_0>0, \qquad i=1,\ldots, d,
$$ 
where $W^{\QQ}= (W^{\QQ,1},\ldots, W^{\QQ,d})$ is a standard Brownian motion in $\RR^d$ with correlation matrix
$(\rho_{i,j})_{1\le i,j\le d}$.
We set $d=100$. A European basket call option is associated with the payoff
$$
g(S^1_T,\ldots, S^d_T) = \left(\sum^d_{i=1} S^i_T -d\cdot K\right)^+.
$$
The results obtained  by Algorithm \ref{algo:solverForNonRecursive} using the parameters  in Table \ref{table:param}
with $\sigma^i = \sigma$ for all $i=1,\ldots, d$, zero correlation, $s^i_0=100$ for all $i=1,\ldots, d$
and $r^i = r=0.01$ are reported in Figure \ref{fig:call} (right).

The distinctive feature of the present example is the high dimension of the vector of risk factors. While the two previous one-dimensional examples mainly served as a validation for the methodology, the present example highlights the ability of the proposed methodology to provide  an accurate numerical approximation in a high-dimensional context. 
For this example, we used the feedforward neural network with two layers and $d+10$ nodes, with a ReLU activation function. 
The approximation parameters used are reported in the caption of Figure \ref{fig:call} (right).
We increase the number of nodes $\nu$ roughly linearly with the dimension $d$, which turned out to be a useful rule-of-thumb for consistent accuracy across dimensions in this case.

For a detailed study of deep learning values of basket derivative (on six underlying asses) from simulated values, not based on BSDEs, see \cite{ferguson2018deeply}.

For the case of the basket call option, we observe that the exposure profile corresponds to the present value of the contract. As a consequence, we obtain a simple method to validate the exposure profile by computing an estimate of the basket call option price by means of a standard Monte Carlo simulation with $10^5$ paths.  We regard this as the `exact' price.  The Monte Carlo price we obtained is $157.99$ with confidence interval $[157.63,  158.34]$. The average values of the expected exposures produced by the deep solver reported in Figure \ref{fig:call} (right) vary  with time between the values $156.98$ and $161.24$ in the positive case, and $0$ and $-2.9824$ in the negative one,  achieving at the terminal time $t=T$ the maximum distance $3.25$ to the Monte Carlo price in the first case and $2.98$ to the exact zero solution in the second case. The accuracy of the time 0 option price is therefore 1bp, and hence of the same order of magnitude as for the single underlying. {The CPU time is $1287$ s}.

\begin{remark}
It is noticeable that the error of DEPE and DENE approximation is relatively low at time zero and eventually increases with time. This is because the time zero value is determined solely by the obtained optimiser for $\xi$, which is decoupled from the harder optimisation problem for $\rho$. The optimal $\xi$ which minimises the idealised objective function without time stepping and sampling error is the  expected payoff, while $\rho$ determines the ANN hedge which minimises the variance. A suboptimal ANN leads to larger hedging errors, and hence increasing DEPE and DENE, as time increases.

One could use this observation to set $\xi$ to be an accurate MC estimator for the option price, and then minimize over $\rho$ only. This by construction gives accurate time zero values for DEPE and DENE, but from our tests (not reported here) leads to similar results  to above for larger $t$.

{ 
 In relation to \eqref{eq:errorYZ}, this shows that the first term on the right-hand side can be made negligible compared to the second term. In these examples, the error of $Y$ and $Z$ are indeed of similar magnitude. This is supported by \eqref{eqn:YZ} in the appendix.
}
\end{remark}

For this product, next, we also perform an xVA calculation with the objective to validate Algorithm \ref{algo:solverForNonRecursive} and Algorithm \ref{algo:solverForXva}
in a case where both are applicable.
To perform this comparison, we need the xVA BSDE to be non-recursive: this can be achieved by assuming that there is a unique risk-free interest rate, so that FVA and ColVA are identically zero, i.e., xVA consists only of the CVA and DVA term. 
The idea is then to compare a Monte Carlo estimate of xVA according to Algorithm \ref{algo:solverForNonRecursive} with the initial value of the BSDE as produced by a full application of Algorithm \ref{algo:solverForXva}.

We assume that the default intensities of the bank and the counterparty are $ \lambda^{C,\QQ}=0.10$ and $\lambda^{B,\QQ}= 0.01$, respectively. For the recovery rates we set $R^C=0.3$ and $R^B=0.4$, while the unique risk-free interest rate is $r=0.01$. 
 Using the same network setting (see again the caption of Figure \ref{fig:call}, right), the Deep xVA Solver produced an xVA estimate of $0.8952$ by means of Algorithm \ref{algo:solverForXva} (CPU = $3098$ s), whereas the estimate produced by Algorithm \ref{algo:solverForNonRecursive} is $0.8947$ with an associated confidence interval  $[0.8927, 0.8968]$ (CPU = $1379$ s).

 {
 As pointed out in Section \ref{sec:riskMeasures}, Algorithm 3 can also be used to compute risk measures for xVAs. In Figure \ref{fig:var} (right) we report the plot of the Value at Risk (VaR, with  $\alpha=0.05$)  of the $\text{xVA}$, i.e. we compute $\text{VaR}_\alpha(\text{XVA}_t)$ taking $L_t=-(\text{XVA}_t-\text{XVA}_0)$, for $t\in[0,T]$, as loss process (simulated paths of the loss are reported in Figure \ref{fig:var} (left)).  Considering the same discretization parameters as the test above the required computational time is  $3425$ s, which confirms that after a single application of Algorithm 3 to solve the xVA BSDE, risk measures can be obtained  at a very low cost just by simulating the associated BSDE trajectories. We acknowledge that the computation of risk measures, which focuses on tails of the distribution of the value process, poses some challenges to our method: we faced numerical instabilities due to the representation of floating point numbers in Python that we addressed by exploiting the fact that both the clean value and the $\text{xVA}$ are homogenous functions of order one with respect to the notional. We multiply the terminal condition of the clean value BSDE by a scaling factor which we later compensate back after the simulation of the paths of the trained model has been performed. We compute the value at risk for every point in the simulation time grid and we observe a smooth curve that converges towards the terminal value at risk. At time $T$ the loss degenerates to $-(\text{XVA}_T - \text{XVA}_0)= \text{XVA}_0 \sim 0.8952$ due to the fact that the $\text{xVA}$ BSDE has a zero terminal condition, hence we have again a test value against which we can compare our estimate given by $\text{VaR}(\text{XVA}_T) = 0.9097$.} 
 \begin{figure}
\includegraphics[width=0.43\textwidth]{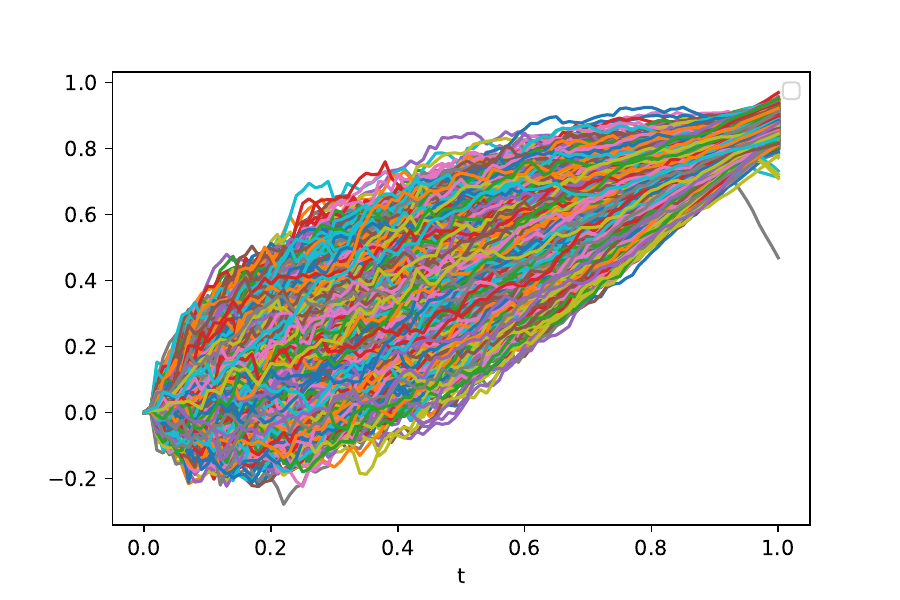}\includegraphics[width=0.43\textwidth]{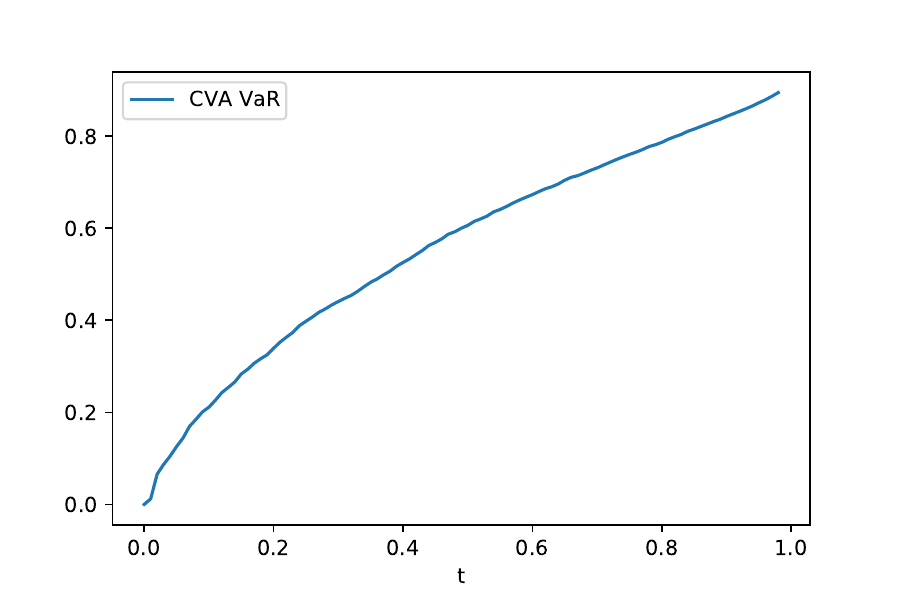}
 \caption{{Simulated paths of the loss process  $L_t = -(\text{XVA}_t - \text{XVA}_0)$ (left) and approximation of the VaR$(\text{XVA}_t)$ for $t\in [0,T]$ (right). }}\label{fig:var}
\end{figure}

\subsection{Recursive FVA computation}

In this section, we provide an FVA calculation that serves as a further validation of Algorithm \ref{algo:solverForXva} for recursive valuation adjustments.
For the sake of illustration, we simplify the framework of \cite{bgo2019} so that we recover the funding equations of \cite{pit10}. More specifically, we assume that there is no default risk, i.e. $\tau^C=\tau^B=+\infty$. We consider the case of a bank trading a forward on a single underlying stock, in line with Example \ref{sec:fwdExample}. We set $r^{c,b}=r^{c,l}=r= 0.02$, and $r^{f,b}=r^{f,l}=0.04$. Due to the different interest rates for funding and collateral, the clean value of the contract is not at par. {For the moment } we assume that the claim is perfectly uncollateralized, i.e.\ $C_t\equiv 0$ $\mathrm d \QQ\otimes \mathrm dt$-a.s. In this case, as first shown in \cite{pit10} and then \cite{bgo2019} among others, one can employ a risk neutral valuation formula where the discount rate is given by the unsecured funding rate $r^f=r^{f,b}=r^{f,l}$. Precisely, we can write the solution of the pricing problem as
\begin{align*}
V_t&=\widehat{V}_t -\overline{\rm FVA}_t, \qquad \text{where} \\
\overline{\rm FVA}_t&=B^{{r}}_t \Excond{\QQ}{\int_t^{T}\frac{(r^{f}_u-r_u)\left(\widehat{V}_u-\overline{\rm FVA}_u\right)}{B^{{r}}_u} \, \mathrm du}{\cF_t}.
\end{align*}

The analytic computation of the clean value of the forward contract at time $t$ yields $\widehat{V}_0^{\mathrm{exact}} = 1.9801$. The claim is however uncollateralized, hence, by applying a risk neutral valuation formula where the discounting rate is now $r^f$, we obtain $V^{\mathrm{exact}}_0 =1.9409$. The difference between the two analytic computations provides us with the exact value of the FVA, i.e.\ $\overline{\rm FVA}^{\mathrm{exact}}_0=0.0392$. For this experiment we apply Algorithm \ref{algo:solverForXva} with the following parameters that are the same both for the estimation of the clean value and the FVA: we use $N = 100$, $L=64$, $P = 2048$ {and $\mathcal I=4000$} . We use two neural networks for the clean value and the FVA both having 2 hidden layers with $d+20$ nodes.
We then apply Algorithm \ref{algo:solverForXva} to the xVA BSDE associated with FVA and obtain an initial value of $\overline{\rm FVA}_0=0.0395$, thus a validation of our proposed numerical procedure.
{We evaluate the performance of the solver in the reconstruction of exposure trajectories computing the average terminal square error on the FVA, i.e.  $\frac1P \sum^P_{p=1} ( \overline{\cX}^{*,(p)}_{N} )^2$ which, for a chosen single run of the algorithm, is found to be $3.36 \times 10^{-5}$  (comparable with the loss function $3.11\times 10^{-5}$ given by the solver ).} \

{To further assess the reliability of the algorithm we test the FVA as a function of the unsecured funding rate $r^f$: as this rate increases, the funding spread has a higher magnitude, meaning that we expect the FVA to increase. Table  \ref{table:FVArate} provides evidence in this regard.}

\begin{table}
\begin{tabular}{|c|c|c|c|}
\hline 
 $r^f$ & $0.04$ & $0.08$ & $0.12$ \\ 
\hline 
Solver & $0.0395$ & $0.1155$ & $0.1897$ \\ 
\hline 
Exact & $0.0392$ & $0.1153$ & $0.1884$ \\ 
\hline 
\end{tabular} 
\caption{{Numerical solution for the FVA for different levels of the unsecured funding rate $r^f$. Parameters used: outer MC paths $P = 2048$, inner MC paths $L =  1024$, { batch size $B=64$}, internal layers $\mathcal{L}-1=  2$, {  nodes of each internal layer} $\nu=d+20 = 21$, iterations $\mathcal I$ = 4000, time steps $N = 100$}. \label{table:FVArate}}
\end{table}

{We also tested the performance of Algorithm 3 with the increasing of the number of risk factors. In Table \ref{table:FVAdim} we report the computational time required by Algorithm 3 for computing the  FVA for a forward contract written on a basket of $d$ underlyings, for different value of $d$. The parameters used in the numerical tests are reported in the caption of the table.}  { We observe that the numerical error is below $1\%$ at least up to dimension $200$. 
}

\textcolor{blue}{
\begin{table}[ht]
\begin{tabular}{|c|c|c|c|c|}
\hline 
$d$ & Deep XVA Sol. & Exact Sol. & {Error} & CPU(s) \\ 
\hline 
 1 &   0.03950 & $0.03920$ &  {0.0003} &605 \\
 10 & $0.39199$ & $0.39209$ & {0.0001} &753\\
25 & $0.97568$ & $0.98023$ &  {0.0046} & 803\\
50 & $1.9439$ & $1.9605$ &  {0.0166} & 960\\
100  & $3.8976$ & $3.9209$ & {0.0233} & 1410 \\
150 &   $5.8603$ &  $5.8813$ & $0.0210$ & 3085\\
200 & 7.8159 & 7.8418 & 0.0258 & 3918\\
\hline 
\end{tabular} 
\caption{{Comparing the computational time  {and the error} for the approximation of the FVA  for a forward written on a basket of $d$ underlyings, with $d=1, 10, 25, 50, 100, 150, 200$. Parameters used: outer MC paths $P = 2048$, inner MC paths $L =  1024$, { batch size $B=64$}, internal layers $\mathcal{L}-1=  2$, { nodes of each internal layer}  $\nu=d+20$, iterations $\mathcal I$ = 4000, time steps $N = 100$. } \label{table:FVAdim}}
\end{table}
 }

\subsection{{Adding collateral}}
A useful feature of our proposed approach consists in the possibility of performing realistic simulations of the collateral account without resorting to
simplifying assumptions.

\subsubsection{Realistic simulation of the collateral account}
We can in fact compute the overall outstanding exposure between the bank and the counterparty by the following steps. Algorithm 1 allows us to simulate paths for all processes $\widehat{V}^m$, $m=1,\ldots,M$. Such paths can then be aggregated so as to produce a simulation of the portfolio process $\widehat{V}=\sum_{m=1}^{M} \widehat{V}^{m}$, that corresponds to the \textit{pre-collateral exposure}. After this, we compute the value of the collateral balance $C$ corresponding to the simulated paths of $\widehat{V}$, which in turn allows us to compute the \textit{post-collateral exposure} process $\widehat{V}-C$ that enters the xVA formulas.

For illustration, we consider $M=1$ and the equity forward from the first example. We introduce a simple example of a collateral agreement where collateral is exchanged between the counterparties at every point in time (a margin call frequency that does not coincide with the simulation  time discretization can of course be treated as well). Collateral is exchanged only in case the pre-collateral exposure is above (below) a receiving (posting) threshold which are both set equal to $5$,   i.e.
$$
C_t:= C(\widehat V_t) = (\widehat V_t -5)^+  - (\widehat V_t +5)^-.
$$
 An illustration for a single path is provided in Figure \ref{fig:collateral1}.
 
 \begin{figure}[h]
\includegraphics[width=0.32\textwidth]{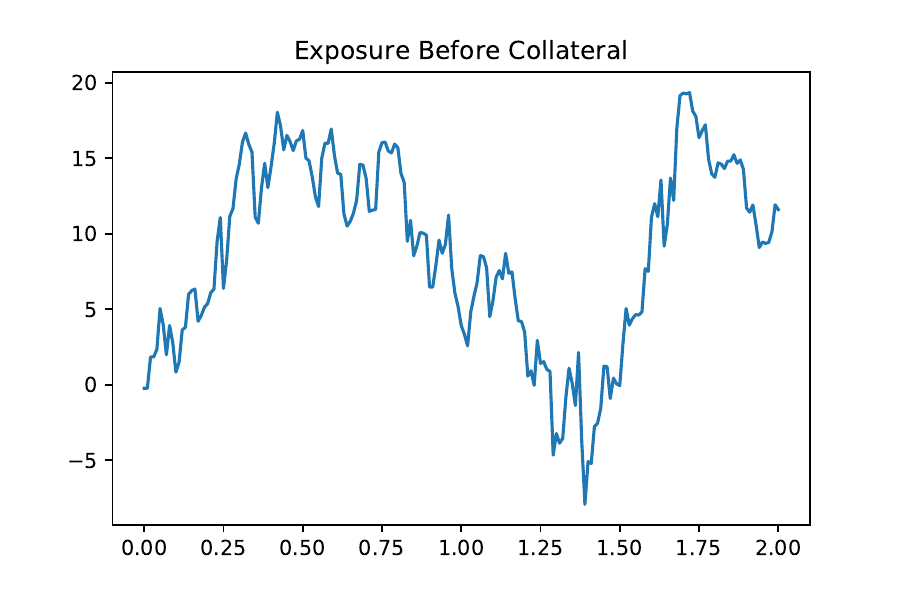}
\includegraphics[width=0.32\textwidth]{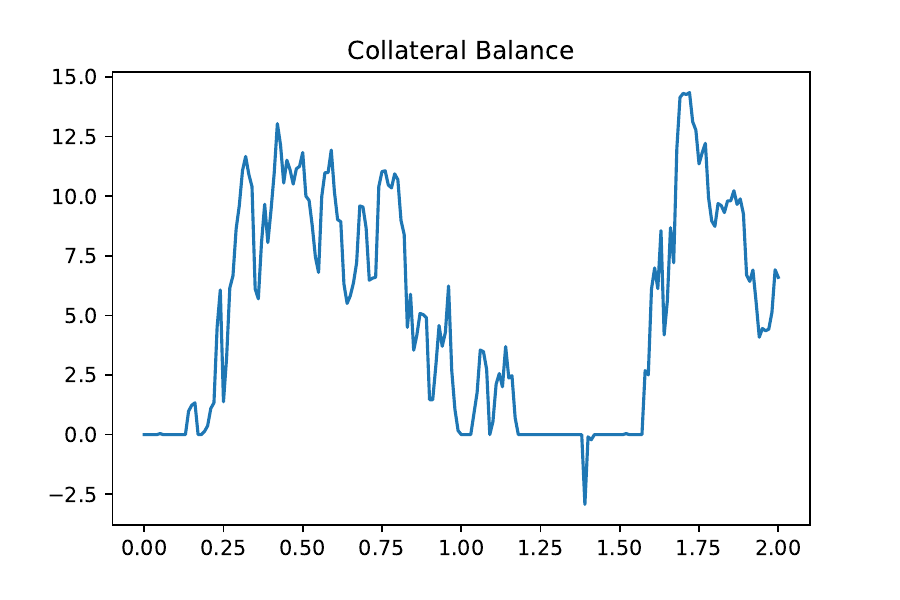}
\includegraphics[width=0.32\textwidth]{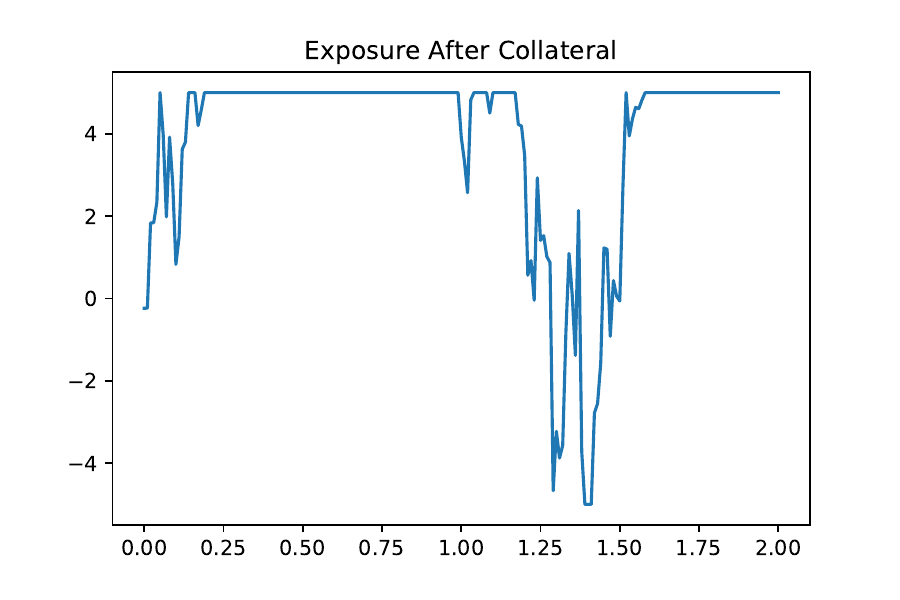}
 \caption{Pathwise simulation of a collateralized exposure. Left: $\widehat{V}$.Middle: $C$. Right: $\widehat{V}-C$. Posting and receiving threshold are $5$ EUR. \label{fig:collateral1}}
\end{figure}

\subsubsection{{Recursive FVA computation in presence of collaterals}}

{We provide further evidence on the algorithm by studying the impact of collateral more closely.
Let us assume that the collateral account is parametrized as follows:
\begin{align*}
C(\widehat{V})=(1-\alpha)[(\widehat{V}-10\alpha)^+- (\widehat{V}+10\alpha)^-], \ \alpha \in [0,1]
\end{align*}
so that the case $\alpha = 0$ corresponds to perfect collateralization and $\alpha=1$ to the uncollateralized case. We compute the FVA of a forward on a basket of underlyings, i.e. we consider the terminal condition
\begin{align*}
g\left(S_{T}^{1}, \ldots, S_{T}^{d}\right)=\left(\sum_{i=1}^{d} S_{T}^{i}-d \cdot K\right), 
\end{align*}
for $K = 100$ where the underlying assets evolve as in Section \ref{subSec:basketCall} and we keep the same choices for the model parameters.  {We make use of Algorithm \ref{algo:solverForXva} assuming $\overline{\rm CVA}=\overline{\rm DVA}=\overline{\rm ColVA}\equiv 0$.}
\textcolor{blue}{
\begin{table}[ht]
\begin{tabular}{|c|c|c|c|c|}
\hline 
$d$ & $\alpha = 0$ & $\alpha = 0.5$ & $\alpha = 1$ & Exact Sol. Uncoll. \\ 
\hline 
$1$ & $4.8408 \times 10^{-5}$ & $0.02222$ & $0.03950$ & $0.03920$ \\
$10$ & $2.8100\times 10^{-5}$ & $0.21164$ & $0.39199$ & $0.39209$ \\
$25$ & $-7.8600\times 10^{-6}$ & $0.48184$ & $0.97568$ & $0.98023$ \\
$50$ & $1.2700\times 10^{-5}$ & $0.72473$ & $1.9439$ & $1.9605$ \\
$100$ & $-6.5900\times 10^{-6}$ & $0.91248$ & $3.8976$ & $3.9209$\\   
\hline 
\end{tabular} 
\caption{{Computation of the FVA for a forward written on a basket of $d$ underlyings. $\alpha = 0$ means full collateralization, where $\alpha = 1.0$ means no collateralization. In the last column we report the exact solution for the no collateral case}.\label{tab:FVAcoll}}
\end{table}
}

{The results are reported in Table \ref{tab:FVAcoll}. We observe, in line with our expectations that in the presence of collateral the FVA becomes numerically negligible. The magnitude of the FVA increases as the level of collateralization decreases. Also we observe that, in the uncollateralized case, the numerical solution provided by the solver closely matches the exact solution for the FVA irrespective of the dimension of the basket of underlying.}

\section{Comparison with {other recent} approaches}

{In this section, we compare our methodology with  other recently developed approaches 
for portfolio-wide xVA computations
from the literature.}

\subsection{Gaussian process regression}

{In \cite{CreDix20}, a Gaussian process regression approach is proposed to perform the computation of derivative portfolio values in the context of (non-recursive) CVA valuation. 
The Gaussian process approximation of the pricing functional is trained on a set of model implied prices, i.e., for a given set of model parameters, a sample of prices is generated for different combinations of contract characteristics (e.g. strike price and maturity). It is clear that, to generate training data, an efficient numerical scheme is needed; for example, for the Heston model, in their Section 3.1, the COS method of \cite{fang2009novel} is employed to generate training data. Once the regression has been trained, efficient pricing and sensitivities computations are possible {even if the portfolio contains a large number of derivatives}. The generation of the training data set appears to be a restrictive aspect of using this methodology,} {especially in the case where individual derivatives are exposed to a large number of risk factors, such as the 100 dimensional basket example we consider. In that case, the generation of gridded price data (the \emph{mark-to-market cube}) over time and all underlying factors is not possible, as remarked in their Section 3.3, and a small number of samples may have to be chosen.\footnote{One could indeed consider using the deep BSDE solver for this scenario generation.} A \emph{divide and conquer} approach allows a significant speed-up through GPU or CPU parallelisation in cases where the portfolio is split up into (asset) classes with a restricted, low number of risk factors.}

{Another limitation emerges when the Gaussian process regression approach is employed to estimate the  Value at Risk of the CVA. To estimate such risk measure, one needs to generate a sample of the random variable  $\text{CVA}_{t+\Delta t} - \text{CVA}_t$, representing the change in value of the CVA between $t$ and $t+\Delta t$, $\Delta t >0$. The application of Gaussian process regression results in a nested loop, which results in a quadratic complexity in terms of the number of simulated scenarios: for $P$ simulated paths, the complexity of the CVA VaR computation of \cite{CreDix20} is $P^2$.
 Instead, with our Algorithm 3,  we directly attack the CVA BSDE and we learn how to simulate trajectories of the CVA, even in high dimension. To estimate the Value at Risk at every possible time horizon within the simulation time grid, we simulate the CVA BSDE with a fixed sample size so that the numerical complexity of our algorithm is lower by one order: we need to simulate only $P$ paths of the clean value and $P$ paths of the CVA BSDE.}


\subsection{{Different deep learning-based BSDE solvers}}

{Within our framework, the Deep BSDE Solver of \cite{ehanjen17} can be directly substituted by other solvers, provided they give approximations to the backward solution and control processes along sample trajectories of the forward process.}
{
\subsubsection*{{The solver developed by \cite{HPW20}}}
This is a regression based scheme for PDEs that runs backward in time making use of dynamic programming arguments. 
The authors study two versions of the algorithm, one where at each point in time two separate neural networks approximate the value function and the control of the BSDE associated to the PDE, and another one where only the value process is approximated by a neural network whereas the control is recovered by automatic differentiation. For each point in time, a neural network is instantiated, and this is common to our methodology. A study of their implementation of the algorithm reveals a problematic point, which is also stated in the text of their paper, that at each point in time, a new Tensorflow session is started and a new computational graph needs to be created, which represents a time consuming operation. Also note that, in order to use their approach for xVA computation, we would need to store each instance of the Tensorflow session on disc so that we can perform the simulation of the exposures after the training is completed. This appears to us as a further bottleneck of the {application of this} methodology {to our setting}.
 A possible advantage that \cite{HPW20} identify for their approach is that only local-in-time optimisation problems have to be solved, compared to the global optimisation problem of the Deep BSDE Solver. The advantage of good initial values for the optimisation available from previous timesteps appears to compete against the error accumulation from projecting the value function onto neural networks in each timestep. }


\subsubsection*{{The approach of \cite{ACHS20}}}
{This framework uses a regression-based algorithm close in spirit to \cite{HPW20}, coupled with Picard iterations for recursive xVA computations and quantile estimation. The authors present a numerical study for the xVA of a swap portfolio with several counterparties and positions exposed to a total of 40 risk factors.
As the single contracts that constitute the portfolio are relatively plain vanilla instruments depending on a low number of risk factors, by a divide and conquer approach  the computation of  single contracts can be split among several GPU/CPU cores. {In principle the regression approach can be extended to cases where individual products depend on a high-dimensional vector of risk factors, such as our high dimensional example for the basket option}.
}

{A comparison of the performance of all these approaches and variants to different practically relevant situations would be of interest, but goes  beyond the scope of this paper.}

\section{Conclusions and extensions}\label{sec:concl}

The proposed xVA algorithm exploits two useful complementary aspects of the Deep BSDE Solver of \cite{ehanjen17}.
First, the formulation as an optimisation problem over a parametrisation of the (Markovian) control of the xVA BSDE, which is carried out by
SDE discretisation and path sampling, directly gives both the hedge ratios in approximate functional form and model-based derivative prices along the sample paths.
This is amenable to the simulation of exposure profiles, the computation of higher-order Greeks by pathwise differentiation, and allows for the computation of funding and margin variation adjustments as well as xVA hedging.
A second aspect of the Deep BSDE Solver is the use of neural networks specifically as parametrisation for the Markovian control. A key advantage 
results from the approximation power of neural networks in high dimensions, which has the potential to make risk management computations on portfolio level feasible.
Moreover, the simple functional form allows standard pathwise sensitivity computations.

Our numerical examples provide a proof of concept, but further systematic testing in realistic application settings is needed.
An additional difficulty arises from the non-linear, non-convex parametric form, which, combined with the large number of parameters, 
leads to challenging optimisation problems.
 The expression power of the ANN and the practicalities of the learning process, are extremely active research areas and further developments of the proposed Deep xVA Solver will be informed by the rapidly developing understanding of neural networks in a broader sense.

The application of our proposed scheme is not restricted to the chosen xVA framework. For example, one could in principle apply our methodology to the balance-sheet based model computed in \cite{ACHS20}. In this case, the xVA computation involves multiple recursive valuations
(illustrated succinctly in \cite[Figure 1]{abbas2018xva}), which can be approached by means of multiple applications of the Deep xVA Solver.

We also emphasise that the Deep xVA Solver can be combined with an existing analytics library: the computation of the mark-to-market cube (i.e., the simulation of all possible scenarios for the clean values over different points in time) represents a classical numerical problem to be solved in order to compute traditional risk figures such as Value-at-Risk or Expected Shortfall (this is often referred to as ``Monte Carlo full revaluation approach'').
Since most products individually depend on a limited number of risk factors, it may be best to use  a traditional numerical scheme, such as a finite difference solver, 
for at least some of the more vanilla products, and then revaluate the products over different Monte Carlo paths by means of a look-up table over the pre-computed numerical solution. This provides an alternative route with respect to our Algorithm \ref{algo:solver} for the simulation of the clean values. However, once we aggregate all mark-to-markets, we end up with an object that depends on a high number of risk factors, so for the computation of xVA our proposed methodology provides a useful tool which allows the recursive computation of valuation adjustments, their hedging strategy, and simulation of collateral.

{Also, let us stress that our Algorithm \ref{algo:solver} returns not only the clean value but also the sensitivies with respect to the forward SDE. The availability of sensitivities is fundamental in order to hedge exposures and also to perform the calculation of initial margin according to the market standard approach (ISDA Simm). In this sense, Algorithm \ref{algo:solver} represents a useful alternative to the above mentioned classical approaches in view of the increased demand of advanced analytics by regulators.}
%
%


\appendix

{ 
\section{Error bounds for $Y$ in terms of $Z$}
\label{app:boundYZ}

We consider the error between the BSDE solution $(Y,Z)$ and its approximation $(Y^{\xi,\rho}, Z^{\rho})$ from the Deep BSDE Solver, satisfying
\begin{eqnarray*}
Y_{t}     & = Y_0 - \int_{0}^{t} h \left(s, X_{s}, Y_{s}, Z_{s}\right) \, \mathrm d s + \int_{0}^{t} Z_{s}^\top \, \mathrm d W^{\QQ}_{s}, \quad t \in[0, T], \\
Y^{\xi,\rho}_{t}     & = \xi - \int_{0}^{t} h \left(s, X_{s}, Y^{\xi,\rho}_{s}, Z^\rho_{s}\right) \, \mathrm d s + \int_{0}^{t} (Z^\rho_{s})^\top \, \mathrm d W^{\QQ}_{s}, \quad t \in[0, T],
\end{eqnarray*}
respectively.
Taking the difference and squaring, by elementary inequalities
\begin{eqnarray*}
|Y_{t} - Y^{\xi,\rho}_{t} |^2 &\le& 3 \left\{
|Y_0 - \xi|^2 + t \int_0^t \left| h \left(s, X_{s}, Y_{s}, Z_{s}\right) - h \left(s, X_{s}, Y^{\xi,\rho}_{s}, Z^\rho_{s}\right)\right|^2  \, \mathrm d s \right. 
\\ && \hspace{6 cm}
\left.
+ \left(
\int_{0}^{t} \left(Z_s -   Z^\rho_{s}\right)^\top \, \mathrm d W^{\QQ}_{s}
\right)^2
\right\}.
\end{eqnarray*}
Taking expectations, using for the second term the Lipschitz continuity of $h$ in $Y$ and $Z$, with constants $L_Y$ and $L_Z$, respectively, and It{\^o} isometry
for the last term,
\begin{eqnarray*}
\EE |Y_{t} - Y^{\xi,\rho}_{t} |^2 &\le& 3 \left\{
\EE |Y_0 - \xi|^2 +
 2 t L_Y^2 \int_0^t \EE \left|Y_{s}- Y^{\xi,\rho}_{s} \right|^2 \, \mathrm d s 
+
( 2 t L_Z^2 + 1) \int_{0}^{t} \EE \left|Z_s -   Z^\rho_{s} \right|^2  \, \mathrm d s
\right\}.
\end{eqnarray*}
 By
 Gronwall's inequality,
\begin{eqnarray*}
\EE |Y_{t} - Y^{\xi,\rho}_{t} |^2 &\le& \left\{
\EE |Y_0 - \xi|^2 +
( 2 t L_Z^2 + 1) \int_{0}^{t} \EE \left|Z_s -   Z^\rho_{s} \right|^2  \, \mathrm d s
\right\}
3 \exp(6 t^2 L_Y^2),
\end{eqnarray*}
which proves \eqref{eq:errorYZ} for a $C$ that only depends on $T$, $L_Y$ and $L_Z$.

Although the inequality is generally not sharp, the order 2 strong error of $Y_t$ is typically close to the $L^2$ error (in $t$ and $\mathbb{Q}$) of $Z$,
as seen numerically in the option pricing examples.
This is supported by the following simple calculation. 
We assume here that $Y_0 = \xi$, justified by the observation that $Y_0$, the option price at time 0, can be accurately found by Monte Carlo estimation.
Then it follows from
\[
Y_t = Y_0 + r \int_0^t Y_s \, \mathrm d s + \int_0^t Z_s \, \mathrm d W^{\QQ}_{s},
\]
using an integrating factor $\exp(-r s)$ and similar steps to above,
\begin{equation}
\label{eqn:YZ}
\EE |Y_{t} - Y^{\xi,\rho}_{t} |^2 = \int_0^t \exp(2 r (t-s)) \, \EE |Z_{s} - Z^{\xi,\rho}_{s} |^2 \, \mathrm d s.
\end{equation}

}

\section{A posteriori error estimates for non-recursive adjustments}\label{app}

The estimates provided in \cite{HanLon18} can be applied as follows to the adjustment computation in Subsection \ref{subsec:nonrec}, Algorithm \ref{algo:solverForNonRecursive}. We assume the existence of some constant $C$ such that 
\begin{equation}\label{eq:errorapp}
\sup_{t\in [0,T]} \EE \Big[ |\widehat V_t  - \widehat V^{\xi, \rho}_t|^2\Big ] \leq C \left( \Delta t + \sum^M_{m=1} \EE\Big[ | g_m(S_{T}) -\widehat{V}^{\, m,\xi_m,\rho_m}_{T}|^2\Big] \right),
\end{equation}
where $\widehat V^{\xi,\rho}_t $ is the ANN approximation associated with parameters $\xi=(\xi_1, \ldots, \xi_M)$ and $\rho =(\rho_1, \ldots, \rho_M)$ (and extended to $[0,T]$ by piecewise constant interpolation) of the clean portfolio value $\widehat V_t$.


Under the assumed conditions on $\Phi$ (uniformly Lipschitz with constant $L_{\Phi}$), one directly obtains  the following estimates 
\begin{align*}
& \bigg| \EE\Big[ \int^T_0  \Phi_t(\widehat V_t) \, \mathrm d t\Big] -  \EE\Big[ \sum^N_{n=0} \eta_n \Phi_{t_n}(\widehat V^{\xi, \rho}_{t_n}) \Big] \bigg| \\
& \leq  \bigg|\EE\bigg[ \int^T_0\Phi_t(\widehat V_t) \, \mathrm d t - \sum^N_{n=0} \eta_n \Phi_{t_n}(\widehat V_{t_n}) \bigg] \bigg| + \bigg| \EE\bigg[\sum^N_{n=0} \eta_n \Big(\Phi_t(\widehat V_{t_n} )  - \Phi_{t_n}(\widehat V^{\xi, \rho}_{t_n}) \Big) \bigg] \bigg|\\
& \leq  \bigg| \int^T_0 \EE\big[\Phi_t(\widehat V_t) \big] \, \mathrm d t - \sum^N_{n=0} \eta_n \EE\big[\Phi_{t_n}(\widehat V_{t_n}) \big] \bigg| + \bigg|\sum^N_{n=0} \eta_n \EE\Big[ \Big(\Phi_t(\widehat V_{t_n} )  - \Phi_{t_n}( \widehat V^{\xi, \rho}_{t_n}) \Big) \Big] \bigg|\\
& \leq Q(\Delta t) + \Big(\sum^N_{n=0} |\eta_n|^2\Big)^{1/2} \Big(\sum^N_{n=0} \left|\EE\left[\Phi_{t_n}(\widehat V_{t_n}) -\Phi_{t_n}(\widehat V^{\xi, \rho}_{t_n})\right]\right|^2\Big)^{1/2}\\
& \leq Q(\Delta t) + L_{\Phi} \Big(\sum^N_{n=0} |\eta_n|^2\Big)^{1/2} \Big(\sum^N_{n=0} \EE\left[|\widehat V_{t_n} -\widehat V^{\xi, \rho}_{t_n}|^2\right]\Big)^{1/2},
\end{align*}
where $Q(\Delta t)$ is the error associated with the quadrature rule for the function $\varphi(t):=\EE[\Phi_t(\widehat V_t)]$. 
The function  $\varphi$ can be proven to be  $1/2$-H\"older continuous. Indeed, 
for $\Phi_t(\widehat{V}_t) = (B^{\tilde{r}}_t)^{-1} \Psi(\widehat{V}_t) \lambda_t^{C,\mathbb{Q}}$ (the CVA case, and similar for DVA),
denoting $\Psi(\widehat{V}_t) = (1-R^C)(\widehat{V}_t-C(\widehat{V}_t))^-$ Lipschitz in $\widehat{V}_t$,
\begin{eqnarray*}
|\phi(t)-\phi(s)| &\le& \\
&& \hspace{-3 cm} \mathbb{E} \left[
\left|(B^{\tilde{r}}_t)^{-1} \Psi(\widehat{V}_t) \left(\lambda_t^{C,\mathbb{Q}}-\lambda_s^{C,\mathbb{Q}}\right)\right|
+
\left|(B^{\tilde{r}}_t)^{-1} \left(\Psi(\widehat{V}_t)-\Psi(\widehat{V}_s)\right) \lambda_s^{C,\mathbb{Q}}\right|
+
\left|\left((B^{\tilde{r}}_t)^{-1} - (B^{\tilde{r}}_s)^{-1}\right) \Psi(\widehat{V}_s) \lambda_s^{C,\mathbb{Q}}\right|
\right] \\
&\le& C \left\{
\mathbb{E}[(\lambda_t^{C,\mathbb{Q}}-\lambda_s^{C,\mathbb{Q}})^2]^{1/2} +
\mathbb{E}[(\widehat{V}_t-\widehat{V}_s)^2]^{1/2} +
\mathbb{E}\Big[\Big(1-\exp\Big(-\int_s^t \tilde{r}_u \, {\rm d} u\Big)\Big)^2\Big]^{1/2}
\right\},
\end{eqnarray*}
for some constant $C$,
using the boundedness of ${r}$, $\lambda^{C,\mathbb{Q}}$, $\lambda^{B,\mathbb{Q}}$, and of $\mathbb{E}[(\Psi(\widehat{V}_t))^2]$.
The first and last term are of order $|t-s|^{1/2}$ by the assumptions made, and it remains to estimate the middle term.

Recalling that $\widehat V_t= \sum^M_{m=1} \widehat V^m_t$ with $\widehat V^m_t$ the solution of the FBSDE \eqref{eq:asset_prox}, \eqref{cleanBSDEeu}, under the regularity assumptions  on the coefficients $\mu$ and $\sigma$ of the forward SDE one gets 
\begin{align*}
\Big|\varphi(t)-\varphi(s)\Big| & \leq C \Big( |t-s|^{1/2} + \sum^M_{m=1} \EE\big[ \big|\widehat V^m_t - \widehat V^m_s\big|^2\big]^{1/2}\Big) \\
& \leq C \Big( |t-s|^{1/2} + \sum^M_{m=1} \EE\big[ |g_m(S_{T})|^2 |t-s| + \int^t_s |\widehat Z^m_u|^2 \, \mathrm d u \big]^{1/2}\Big)\\
&  \leq C \Big( |t-s|^{1/2} + |t-s|^{1/2}  \sum^M_{m=1} \EE\big[ |g_m(S_{T})|^2 + 1 + \sup_{u\in [s,t]} |S_u|^2 \big]^{1/2}\Big) \\
&  \leq C |t-s|^{1/2}.
\end{align*}
To obtain the estimates for the increment of the BSDE solution and of the control in terms of the forward process, in the second and third line, respectively, 
we can apply \cite[Lemma 2.4, (2.11)]{Zhang04} and \cite[Theorem 5.2.2(i)]{Zhang_book} to the equivalent BSDE
\begin{eqnarray}
\label{trans}
\qquad
{\rm d} \widetilde{V}_t^m = \widetilde{Z}_t^m \, {\rm d} W_t^{\mathbb{Q}},
\quad \widetilde{V}_T^m = g_m(S_T),
\qquad \text{where} \quad \widetilde{V}_t^m = \widehat{V}_t^m B^r_T/B^r_t,
\quad \widetilde{Z}_t^m = \widehat{Z}_t^m B^r_T/B^r_t.
\end{eqnarray}
Above and in the following, we do not keep track of constants and $C$ denotes any non-negative constant depending only on $T, M, s_0$ and the regularity constants of the coefficients.

Then, if we consider the rectangle quadrature rule we get 
\begin{equation}\label{eq:quad}
Q(\Delta t)\leq C \Delta t^{1/2} .
\end{equation}
Therefore, observing that 
$$
\Big(\sum^N_{n=0} |\eta_n|^2\Big)^{1/2} \leq \Delta t N^{1/2}
$$
and using \eqref{eq:errorapp} and \eqref{eq:quad}, one has 
\begin{align*}
&\bigg| \EE\Big[ \int^T_0  \Phi(t, \widehat V_t) \, \mathrm d t\Big] -  \EE\Big[ \sum^{N-1}_{n=0} \Delta t \Phi(t_n, \widehat V^{\xi, \rho}_{t_n}) \Big] \bigg|\\
 &\quad  \leq C \Delta t^{1/2}  + C \Delta t N^{1/2} N^{1/2}\Big( \sup_{n=0,\ldots,N-1} \EE\left[|\widehat V_{t_n} -\widehat V^{\xi, \rho}_{t_n}|^2\right]\Big)^{1/2}\\
&\quad  \leq C \Delta t^{1/2}  + T C \Big( \Delta t + \sum^M_{m=1} \EE\Big[ | g_m(S_{T}) -\widehat{V}^{\, m,\xi_m,\rho_m}_{T}|^2\Big] \Big)^{1/2},
\end{align*}
from which the claim \eqref{eqn:nonrec_error}
follows  just taking $\xi=(\xi_1^*, \ldots, \xi^*_M)$ and $\rho= (\rho_1^*, \ldots, \rho^*_M)$.

\section{Hyperparameters tuning}
{
The aim of this section is to test the response of our algorithm with respect to changes in the hyperparameters. In particular, we will study the behaviour of the solution and of the loss function with respect to the variation of the number of iterations $\cI$, the number of hidden layers $\cL-1$ and the value of the learning rate used in the stochastic gradient descent algorithm.
\\
{We focus on the cases of the forward contract and the call option for which we can provide exact solutions as benchmarks.}
Table \ref{app:iterations} shows the impact of increasing the number of iterations on the loss function. We observe a substantial reduction of the loss  as the number of iterations increases from $100$ to $10000$.   {We point out that the increse in the number of iterations has to be coupled with a suitable schedule of the learning rate.} The computational time of course increases, showing that the reduction of the loss comes at a non-negligible cost in terms of computational time. In Figures \ref{app:iterationsFigure} and  \ref{app:iterationsFigure_call} we can qualitatively observe that the exposure profile produced by the solver is in good agreement with the analytical one even for  {$\cI=500$, a relatively } low number of iterations. {In the numerical tests in Section \ref{sec:numerics} we used $\cI=4000$.}

\smallskip
\ \ 
\begin{table}[h!]
\begin{tabular}{|c|c|c|}
\hline 
$\cI$ & loss & CPU(s)\\ 
\hline 
100 & 4.6802e+02 & 103 \\
200 & 2.1080e+01 & 135 \\
500 & 4.3487e+00 & 163 \\
1000 & 2.9029e+00 & 283\\
4000 & 1.6640e-01 & 503\\
10000 & { 3.2899e-02}  & 765  \\ 
 { 40000 } &  {  6.7284e-03} &  {  3194} \\
 { 60000} &  { 5.5881e-03} &  { 4874} \\
\hline 
\end{tabular}
\qquad\qquad
\begin{tabular}{|c|c|c|}
\hline 
$\cI$ & loss & CPU (s)\\ 
\hline 
100 & 2.8109e+02 & 144 \\
200 & 5.7360e+01 & 157 \\
500 & 1.2659e+01 & 241 \\
1000 &  1.1770e+01 & 284\\
4000 & 7.7790e-01 & 543 \\
10000 &  { 5.6281e-01} & 798\\
 { 40000} &  { 4.9823e-01} &  { 3267} \\
 { 60000} &   { 4.0948e-01} &  { 5124}  \\
\hline 
\end{tabular}
\caption{\label{app:iterations}{Variation in the loss function for different values of the number $\cI$ of iterations for the forward contract (left) and the call option (right). Parameters used: outer MC paths $P = 2048$, inner MC paths $L =  1024$, { batch size $B=64$}, internal layers $\mathcal{L}-1 = 2$, { nodes of each internal layer} $\nu = d+20 = 21$, time steps $N = 200$}}
\end{table}


\begin{center}
 \begin{figure}[h]
 \begin{minipage}[h]{0.4\textwidth}
\centering
\includegraphics[width=1\textwidth]{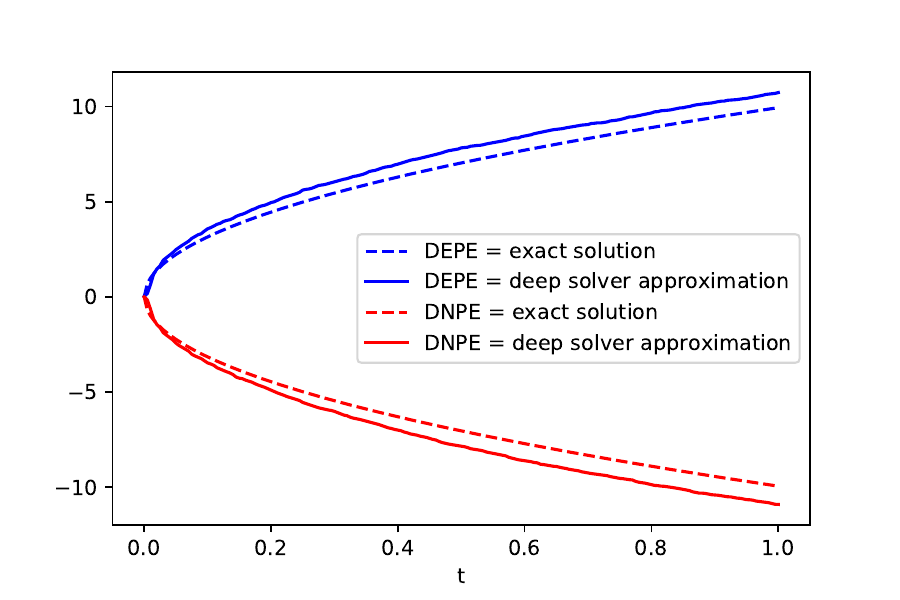}\\{$\cI=200$}
\end{minipage}
 \begin{minipage}{0.4\textwidth}
\centering
\includegraphics[width=1\textwidth]{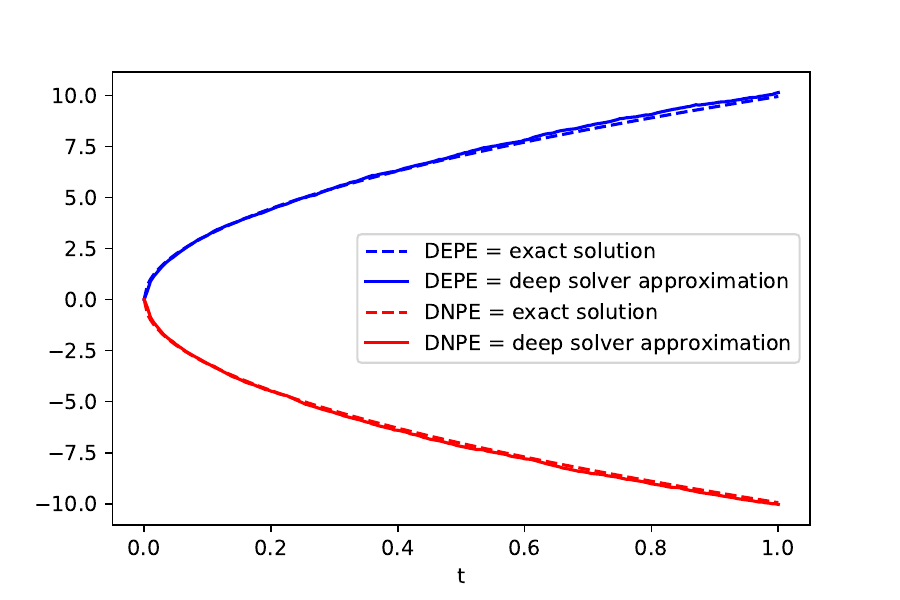}\\{$\cI=500$}
\end{minipage}\\
 \begin{minipage}{0.4\textwidth}
\centering
\includegraphics[width=1\textwidth]{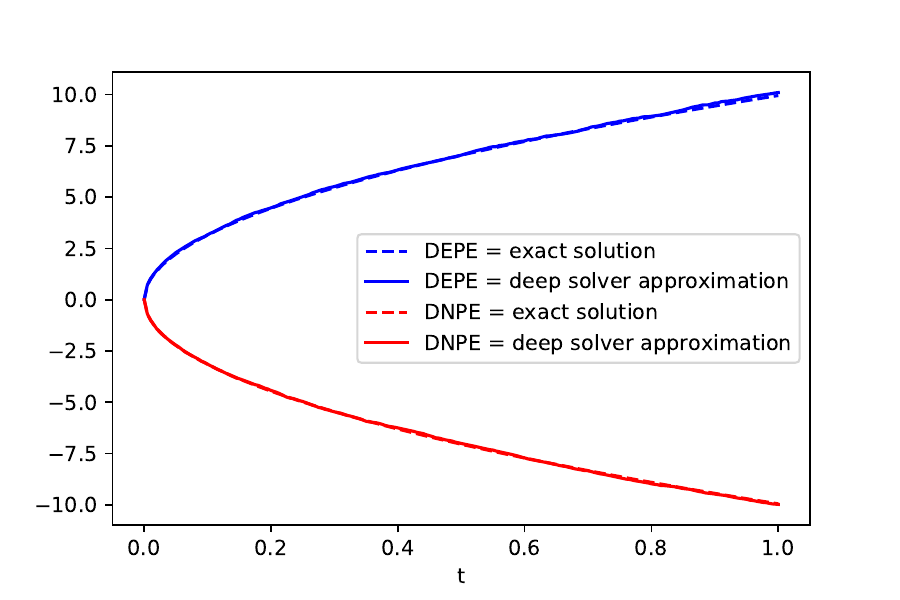}\\{$\cI=4000$}
\end{minipage}
 \begin{minipage}{0.4\textwidth}
\centering
\includegraphics[width=1\textwidth]{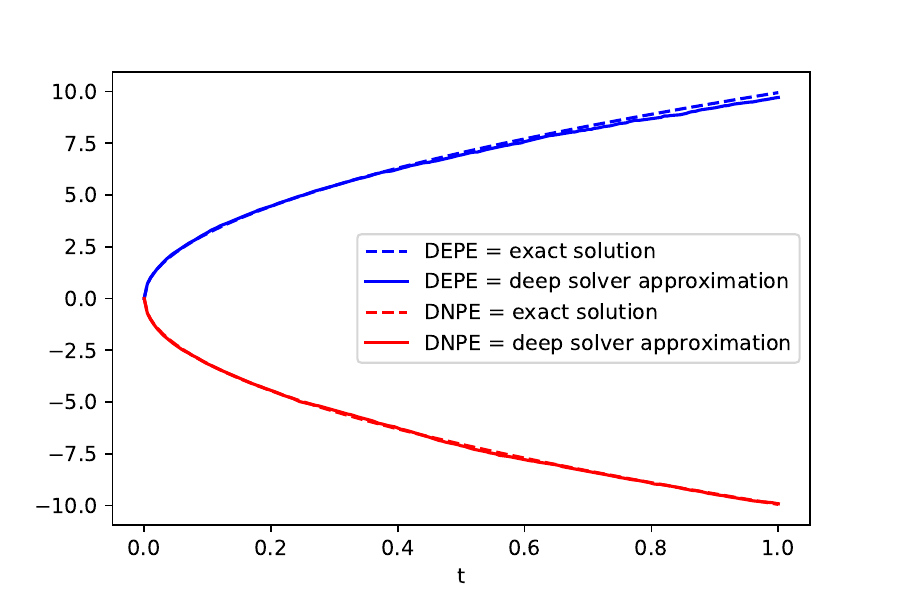}\\{$\cI=10000$}
\end{minipage}%
\caption{Exposure profile for the forward contract for different numbers of iterations. Parameters used: outer MC paths $P = 10000$, inner MC paths $L =  1024$, {  batch size $B=64$}, internal layers $\mathcal{L}-1 = 2$, {  nodes of each internal layer $\nu = d+20 = 21$}, time steps $N = 200$ \label{app:iterationsFigure}}
\end{figure}
\end{center}

\begin{center}
 \begin{figure}[h]
 \begin{minipage}[h]{0.4\textwidth}
\centering
\includegraphics[width=1\textwidth]{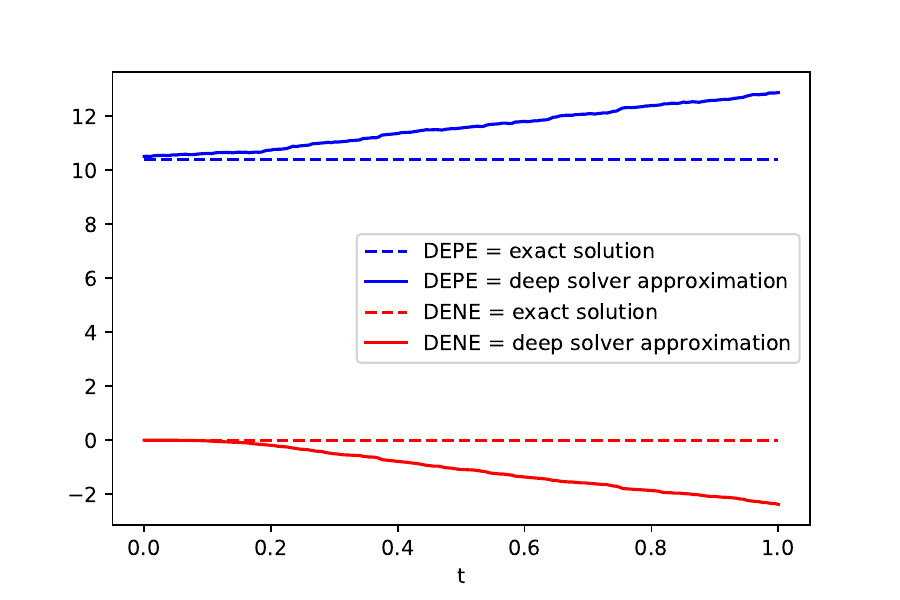}\\{$\cI=200$}
\end{minipage}
 \begin{minipage}{0.4\textwidth}
\centering
\includegraphics[width=1\textwidth]{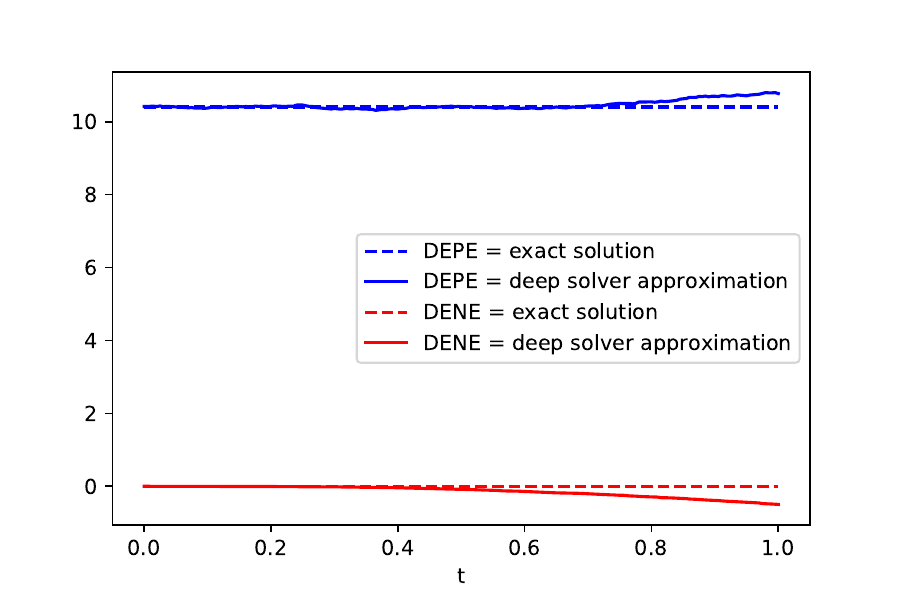}\\{$\cI=500$}
\end{minipage}\\
 \begin{minipage}{0.4\textwidth}
\centering
\includegraphics[width=1\textwidth]{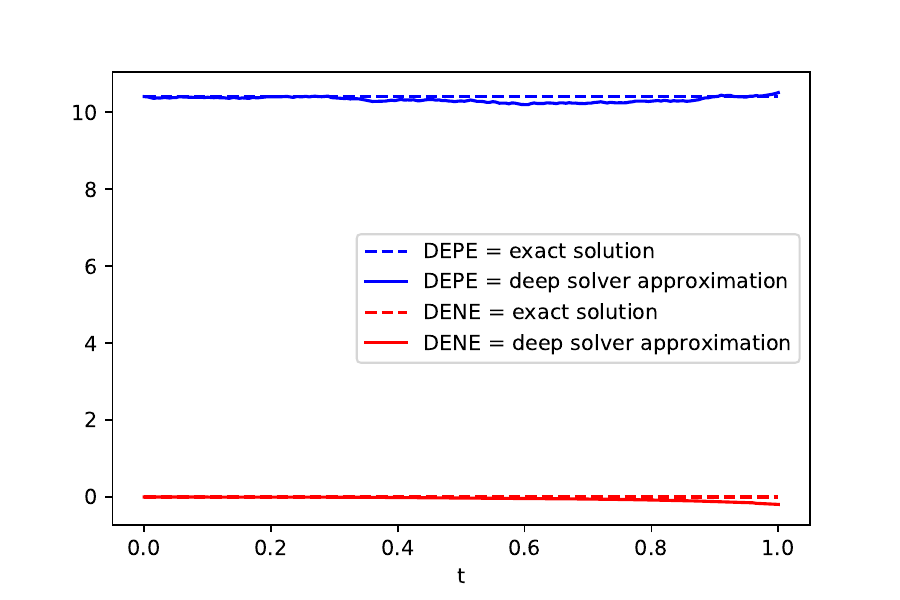}\\{$\cI=4000$}
\end{minipage}
 \begin{minipage}{0.4\textwidth}
\centering
\includegraphics[width=1\textwidth]{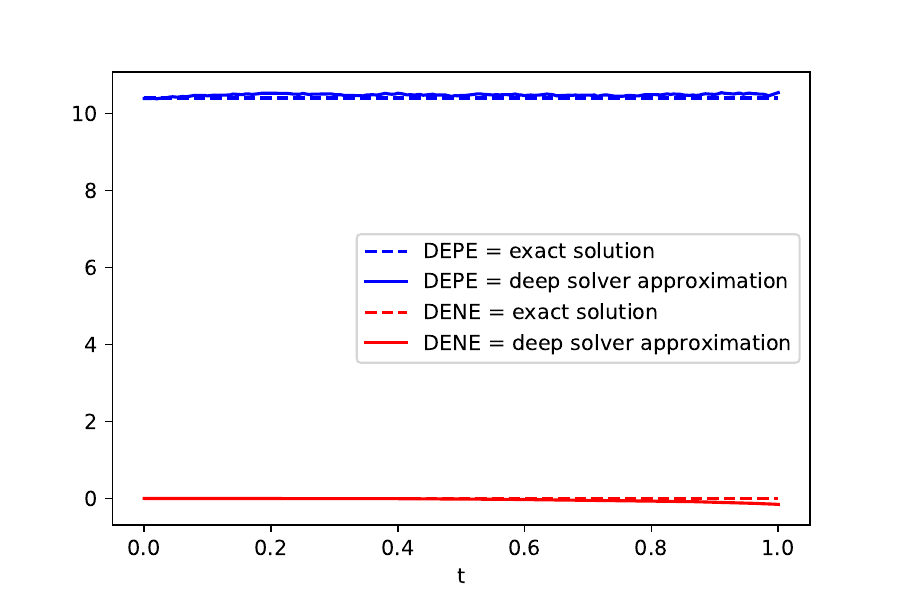}\\{$\cI=10000$}
\end{minipage}%
\caption{Exposure profile for the call option for different numbers of iterations. Parameters used: outer MC paths $P = 10000$, inner MC paths $L =  1024$, { batch size $B=64$}, internal layers $\mathcal{L}-1 = 2$, {  nodes of each internal layer} $\nu = d+20 = 21$, time steps $N = 200$ \label{app:iterationsFigure_call}}
\end{figure}
\end{center}

The next investigations involve the architecture of the network. {  We firstly progressively increased the number of hidden layers from $1$ up to $8$. We performed the  test for  the call option since in  the case of the forward contract a  single layer is sufficient to provide a very good fit. The results are reported in Table \ref{tab:layersCall}. We experienced a moderate reduction of the loss when increasing the number of hidden layers beyond $2$. The loss can be further reduced by coupling the increase of the number of layers with a higher number of iterations (see right-hand table).  Secondly, we increased the number of nodes from $d+20=21$ to $d+200=201$, fixing $2$ hidden layers. The results reported in Table \ref{tab:nodesCall} show again a small reduction of the loss. We tested two different numbers of iterations ($\mathcal I=4000$ and $\mathcal I=40000$, top row) and batch sizes ($B=64$ and $B=256$, bottom row), obtaining a further loss reduction from increasing $\mathcal I$ and $B$.
 }

\begin{table}[h!]
\begin{tabular}{|c|c|c|}
\hline 
$\cL-1$ & loss & CPU (s)\\ 
\hline 
1 &  2.5790e+00  & 356 \\
2 &  7.7790e-01  & 543 \\
4 & {  6.3602e-01 }  &  781\\
8 &  {  6.8549e-01} & 1486 \\
\hline 
\end{tabular}
\qquad \qquad
{ 
\begin{tabular}{|c|c|c|}
\hline 
$\cL-1$ & loss & CPU (s)\\ 
\hline 
1 &   2.1964e+00 & 2053 \\
2 &  4.9823e-01  &  3267\\
4 &  4.8404e-01 &  5388\\
8 & 4.2942e-01 & 10434 \\
\hline 
\end{tabular}
}
\caption{Call option. Variation of the  loss function for different number $\cL-1$ of hidden layers. Parameters used: outer MC paths $P = 2048$, inner MC paths $L =  1024$, { batch size $B=64$},  {  nodes of each internal layer} $\nu = d+20 = 21$, time steps $N = 200$,  iterations $\mathcal{I}= 4000$ {  (left) and $\mathcal I= 40000$ (right).} }\label{tab:layersCall}
\end{table}

\begin{table}[h!]
{ 
\begin{tabular}{|c|c|c|}
\hline 
 $\nu$ & loss & CPU (s)\\ 
\hline 
$21 $ & 7.7790e-01 & 543\\
 $201$ &  6.4578e-01 & 1741\\
\hline 
\end{tabular} \quad\quad 
\begin{tabular}{|c|c|c|}
\hline 
$\nu$ & loss & CPU (s)\\ 
\hline 
$21 $ & 4.9823e-01  & 3267 \\
$201$ &   4.2650e-01 & 10425 \\
\hline 
\end{tabular}
\vspace{0.2cm}\\
$B= 64$, $\mathcal I= 4000$\qquad \hspace{2cm} $B= 64$, $\mathcal I= 40000$\vspace{0.8cm}\\
\begin{tabular}{|c|c|c|}
\hline 
$\nu$ & loss & CPU (s)\\ 
\hline 
$21 $ & 5.6281e-01 & 798\\
$201$ &  5.0817e-01  & 2922 \\
\hline 
\end{tabular} \quad\quad 
\begin{tabular}{|c|c|c|}
\hline 
$\nu$ & loss & CPU (s)\\ 
\hline 
$21 $ & 4.0739e-01 & 1481 \\
$201$ &  3.5423e-01  & 9654 \\
\hline 
\end{tabular}
\vspace{0.2cm}\\
$B= 64$, $\mathcal I= 10000$\quad \hspace{2cm} $B= 256$, $\mathcal I= 10000$\vspace{0.2cm}\\
\caption{Call option. Variation of the  loss function for different number of nodes $\nu$. Parameters used: outer MC paths $P = 2048$, inner MC paths $L =  1024$, time steps $N = 200$, internal layers $\mathcal{L}-1 = 2$. {  Top row: batch size $B=64$, iterations $\mathcal{I}= 4000$ (left) and $\mathcal{I}= 40000$ (right). Bottom row:  batch size $B=64$ (left) and $B=256$ (right), iterations $\mathcal{I}= 10000$} . }\label{tab:nodesCall}
}
\end{table}

\smallskip
\ \ 
We then study the impact of changing the learning rate. Remember that the learning rate  represents the step size of the parameters updates during the training procedure. In our algorithm, we employ a so-called \textit{step decay}, meaning that the learning rate is piecewise constant over the number of iterations. { In the next experiments, we use a learning rate schedule  of the form $[lr1,lr2]$, with two values $lr1, lr2$, where $lr1$ is used for the first half of the  iterations, whereas $lr2$ is used for the remaining iterations}.  {From the tests reported in Table \ref{tab:LRforward1} and  Table \ref{tab:LRforward},  we deduce that the choice of a too high  learning rate may negatively affect the results, while for low learning rates a higher number of iterations is required to ensure sufficient training of the network. In the numerical tests in Section \ref{sec:numerics} we have chosen the schedule $[5e-2,5e-3]$, which was generally observed to provide a good fit}  {both for a high and a low number  of iterations}.}
\\
 {Finally, Figure \ref{fig:time_batches} shows the convergence in terms of time discretization and batch size for the call option. We observe that increasing the batch size improves the convergence behavior. When the batch size reaches $256$, the error levels off so that a further increase does not give a significant improvement, keeping in mind also the associated increase in the execution time of the algorithm. The right panel shows the effect of refining the time discretization. 
We observe that, as we increase the number of time steps up to  $200$, the numerical error is reduced, while further increases do not bring further reductions of the loss. This is possibly due to the significant increase in the number of neural network parameters to be estimated (one network per timestep), while $\mathcal I$ and $L$ are not modified in these tests.

\begin{table}[h!]
\begin{tabular}{|c|c|c|}
\hline 
l.r. & loss & CPU(s)\\ 
\hline 
$\text{[1e-3,1e-4]}$   &   4.7649e+01  &  469\\
$\text{[5e-3,5e-4]}$   &  1.5160e+00   & 445\\
$\text{[1e-2, 1e-3]}$  &  5.6159e-01   & 476 \\
$\text{[5e-2,5e-3]}$   & 1.6640e-01 & 503  \\
$\text{[1e-1,1e-2]}$   &  1.7459e-01 & 423 \\
$\text{[5e-1,5e-2]}$   & 1.3840e+01 & 434\\
\hline 
\end{tabular}
\qquad\qquad
\begin{tabular}{|c|c|c|}
\hline 
l.r. & loss & CPU (s)\\ 
\hline 
$\text{[1e-3,1e-4]}$   & 1.1080e+01    & 392  \\
$\text{[5e-3,5e-4]}$   &  8.2645e-01   & 399\\
$\text{[1e-2, 1e-3]}$  &   6.1086e-01  & 462  \\
$\text{[5e-2,5e-3]}$   & 7.7790e-01 & 543  \\
$\text{[1e-1,1e-2]}$   & 1.0811e+00  &  498 \\
$\text{[5e-1,5e-2]}$   &  1.0183e+01 & 420\\
\hline 
\end{tabular}
\caption{{Effects of changes of the learning rate for the forward contract (left) and the call option (right). Parameters used: outer MC paths $P = 2048$, inner MC paths $L =  1024$, { batch size $B=64$}, iterations $\mathcal{I}= 4000$, number of hidden layers $\cL-1=2$, {  nodes of each internal layer} $\nu = d+20 = 21$, time steps $N = 200$. The first value of the learning rate is used for the first $2000$ iterations.}}\label{tab:LRforward1}
\end{table}

 {
\begin{table}[h!]
{ 
\begin{tabular}{|c|c|c|}
\hline 
l.r. & loss & CPU(s)\\ 
\hline 
$\text{[1e-3,1e-4]}$   &   1.2445e+00   & 1523  \\
$\text{[5e-3,5e-4]}$   &  4.3438e-02   &  1539\\
$\text{[1e-2, 1e-3]}$  &   1.0684e-01  & 1565\\
$\text{[5e-2,5e-3]}$   &  7.8567e-02 &  1516 \\
$\text{[1e-1,1e-2]}$   &    1.6278e-01& 1666  \\ 
$\text{[5e-1,5e-2]}$   &   1.4697e+01& 1691\\
\hline 
\end{tabular}
\qquad\qquad
\begin{tabular}{|c|c|c|}
\hline 
l.r. & loss & CPU (s)\\ 
\hline 
$\text{[1e-3,1e-4]}$   & 6.4950e-01    & 1769   \\
$\text{[5e-3,5e-4]}$   &    5.6589e-01 & 1815 \\
$\text{[1e-2, 1e-3]}$  &  6.0227e-01  &  1903 \\
$\text{[5e-2,5e-3]}$   & 6.6738e-01 &  1767 \\
$\text{[1e-1,1e-2]}$   &   8.9118e-01  &  1732  \\
$\text{[5e-1,5e-2]}$   & 5.1818e+01  & 1824\\
\hline 
\end{tabular}
\caption{{ {Effects of changes of the learning rate for the forward contract (left) and the call option (right). Parameters used: outer MC paths $P = 2048$, inner MC paths $L =  1024$, {  batch size $B=64$}, iterations $\mathcal{I}= 20000$, number of hidden layers $\cL-1=2$, {  nodes of each internal layer} $\nu = d+20 = 21$, time steps $N = 200$. The first value of the learning rate is used for the first $10000$ iterations.}}}\label{tab:LRforward}
}
\end{table}
}

 \begin{figure}[h]
\includegraphics[width=0.32\textwidth]{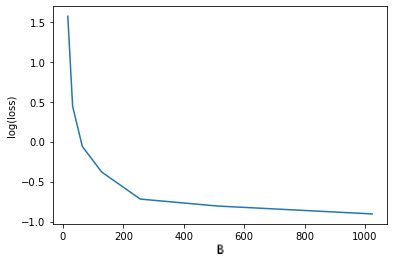}
\includegraphics[width=0.32\textwidth]{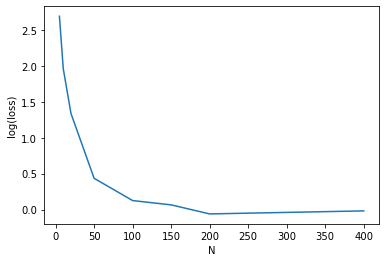}
 \caption{ {Call option. Left: convergence with respect to batch size $B$, parameters used:  iterations $\mathcal{I}= 4000$, time steps $N=200$. Right:  convergence with respect to the number of time steps $N$, parameters used:  iterations $\mathcal{I}= 4000$, batch size $B=64$. }. \label{fig:time_batches}}
\end{figure}

}

\bigskip

{\bf Acknowledgements:} The authors thank Chang Jiang for valuable contributions to the code during his MSc in Mathematical and Computational Finance at Oxford University. We are also grateful to St{\'e}phane Cr{\'e}pey for instructive comments on xVA modelling generally and our framework specifically. The authors thank also Martin Hutzenthaler for useful comments on a previous version of this work.

\bibliographystyle{apa}
\bibliography{biblio}

\end{document}